\documentclass[aps,prb,reprint,longbibliography,superscriptaddress,floatfix]{revtex4-2}

\usepackage{mathrsfs}
\usepackage{amsmath,gensymb}
\usepackage{amsfonts}
\usepackage{amssymb}
\usepackage{amsthm}
\usepackage{graphicx}
\graphicspath{{figures/}}
\usepackage{natbib}
\usepackage{color}
\usepackage{xcolor}
\usepackage{hyperref}
\hypersetup{hidelinks}
\usepackage{bm}
\usepackage[caption=false]{subfig}
\usepackage{verbatim}

\begin{document}
\title{Vortex Core-to-Soliton Conversion in a Mesoscopic Superconducting Grain}

\author{A.~A.~Golubov}
\affiliation{Moscow Institute of Physics and Technology, 141700 Dolgoprudny, Russia}
\affiliation{HSE University, 101000 Moscow, Russia}

\author{M.~Yu.~Kupriyanov}
\affiliation{Moscow Institute of Physics and Technology, 141700 Dolgoprudny, Russia}
\affiliation{Skobeltsyn Institute of Nuclear Physics, Lomonosov Moscow State University,
Moscow 119991, Russian Federation}

\author{M.~M.~Khapaev}
\affiliation{Department of Numerical Methods,
Faculty of Computational Mathematics and Cybernetics, Lomonosov Moscow State University,
Moscow 119991, Russian Federation}

\author{A.~S.~Mel'nikov}
\affiliation{Moscow Institute of Physics and Technology, 141700 Dolgoprudny, Russia}
\affiliation{Institute for Physics of Microstructures, Russian
Academy of Sciences, 603950 Nizhny Novgorod, GSP-105, Russia}

\author{V.~S.~Stolyarov}
\affiliation{Moscow Institute of Physics and Technology, 141700 Dolgoprudny, Russia}

\begin{abstract}
We study an Abrikosov vortex trapped in a circular superconducting grain of
radius $R_0$. The grain is coupled to a surrounding film of the same
material through a closed resistive interface with dimensionless resistance
$\Gamma$. We describe the system within two-dimensional Ginzburg-Landau theory
near $T_c$, with interface conditions derived from Usadel theory. We then
follow how the vortex changes as the coupling weakens, that is, as $\Gamma$
increases. For an isolated vortex in the zero-field limit, we find two
parameter-free results. First, for a nearly transparent interface
($\Gamma\to0$), the centered vortex is stable against small displacements
only if $r_0=R_0/\xi_{\mathrm{GL}}(T)$ is smaller than $r_0^*\simeq0.85$,
where $\xi_{\mathrm{GL}}(T)$ is the Ginzburg--Landau coherence length.
Second, in small grains ($r_0\to0$), the centered vortex loses stability at
$\Gamma_{\mathrm{tr}}\simeq4/(3r_0)$. At larger $\Gamma$, the vortex core
moves continuously toward the interface. When the core reaches it, the
order parameter vanishes only on the grain side, and the core is absorbed.
The winding inside the grain then changes from one to zero, while the fields
and the free energy evolve continuously. The resulting state is an
interfacial relative-phase soliton without a point core. The phase still
winds by $2\pi$ around the grain but not inside it, and the missing winding
is carried by a $2\pi$ kink of the phase difference across the interface.
This state is the Josephson-vortex configuration of a closed (Corbino) weak
link, but here it forms from a trapped Abrikosov vortex. At a finite magnetic field, 
neighboring vortices confine the trapped one,
and we model this effect with a circular cell that has a fixed outer
boundary. In this cell, the onset of vortex displacement changes from
continuous to first order at a tricritical point as the grain radius
increases. For still larger grains, the vortex switches directly from the
centered state to the soliton, with hysteresis. These results show that a
grain boundary can change not only the position but also the core structure
of a vortex.
\end{abstract}

\maketitle

\section{Introduction}
\label{sec:introduction}

An Abrikosov vortex is characterized by a $2\pi$ phase winding around a point at which the superconducting order parameter vanishes. The amplitude returns to its bulk value on the scale of the coherence length \cite{Abrikosov1957,Tinkham1996}. Although this point-core picture is standard, boundaries and weak links can deform the core and modify the vortex structure.

Motivated by this possibility, we study a superconducting grain bounded by a resistive interface. The central question is whether the point zero associated with a vortex can disappear at the interface while the phase winding on a distant contour remains $2\pi$. Such a process would show that the interface can change the core structure as well as the vortex position.

Ginzburg--Landau (GL) studies of mesoscopic disks, cylinders, and polygonal samples show that confinement can reorganize vortex matter. Multivortex, giant-vortex, and symmetry-broken states can exchange stability through continuous or first-order transitions \cite{Schweigert1998,Baelus2002,Chibotaru2000}. Thermodynamic crossings can occur before a metastable state disappears \cite{Baelus2001}. The minimum sample size, surface pair breaking, and holes also affect the vortex configurations \cite{Zharkov2001,Pogosov2001,Pogosov2002Surface,Berdiyorov2004}. A vortex that approaches an artificial defect can develop an elongated region of suppressed order parameter that reaches the boundary \cite{Priour2003}. Bespalov and Mel'nikov derived analytical GL pinning potentials for small insulating cavities, including relaxation of the surrounding order parameter \cite{BespalovMelnikov2013}. In our system, a closed resistive interface couples two superconducting regions with independent complex order parameters. Therefore, both the amplitude jump and the relative phase can influence the vortex structure.

A second line of related work concerns vortices at weak links and grain boundaries. Low-angle grain boundaries support Abrikosov--Josephson vortices, whose cores extend along the weak boundary \cite{Gurevich2002Theory,Gurevich2002}. The crossover between Abrikosov and Josephson vortices has been discussed for a long time \cite{likharev1975a,likharev1975b}. Time-dependent GL calculations for lateral $S/S'/S$ weak links found that a current drives the vortex toward a Josephson-like phase object \cite{Carapella2016}. Microscopic calculations showed that the junction transparency tunes this crossover, and that a vortex pinned at the junction can lose its normal core \cite{Samokhvalov2020}; a diffusive normal region can strongly expand the vortex core \cite{Stolyarov2018}. In Ref.~\cite{Samokhvalov2020}, the vortex sits on an open planar defect, and the paper studies its quasiparticle spectrum at a fixed position. Here the interface is closed, and the core can move. The question is then topological: can the winding inside the grain change while the winding on a contour outside the grain stays equal to one?

We denote the azimuthal angle around the grain center by $\varphi$. The state that we find below is an interfacial relative-phase soliton. It has a $2\pi$ winding of the relative phase $\delta(\varphi)$ between the outer and the inner condensate, and no point zero of the order parameter in either region. In the calculated states, the phase variation is localized along the interface. Relative-phase solitons also exist in multiband superconductors when the windings of the components differ \cite{Kuplevakhsky2011,Vakaryuk2012,Samokhin2012}. In addition, mesoscopic two-gap disks support asymmetric and fractional-vortex states with unequal component windings \cite{Geurts2010,Pina2012}.

Closed weak links are the closest geometric analog of our grain interface. Clem obtained zero-current Josephson-vortex phase configurations for planar Corbino weak links \cite{Clem2010}. Later, time-dependent GL studies considered the current-driven nucleation of vortices and how vortices are guided into Corbino junctions \cite{Okugawa2022}. Radius bottlenecks in superconducting tubes can localize fluxoid solitons \cite{Kokkeler2026}, and a fluxoid mismatch in a Corbino junction can produce a phase soliton that binds Andreev states \cite{SanJose2026}. In such a closed junction, a winding difference of one between the inner and outer electrodes is carried by a $2\pi$ kink of the phase difference along the junction, the Josephson-vortex configuration of the Corbino geometry \cite{Clem2010}. Our interfacial relative-phase soliton has exactly this topology. A similar competition between an interfacial Josephson vortex and a centered vortex was found for Ru inclusions in Sr$_2$RuO$_4$ \cite{Kaneyasu2010}. In these systems, the winding mismatch is set from outside. It comes from the boundary conditions~\cite{Clem2010}, from the electrode geometry and an axial flux~\cite{Kokkeler2026,SanJose2026}, or from the chiral order parameter~\cite{Kaneyasu2010}. In our problem, only a unit winding far away is imposed. The mismatch appears when the interface absorbs the point core of a trapped Abrikosov vortex.

In this work, we analyze how this state can form from an Abrikosov vortex trapped in the grain, as its singular point core is absorbed by the interface. The order-parameter amplitudes on the two sides of the interface are obtained self-consistently, and the only imposed topological constraint is a unit winding on a remote contour. When the displaced core reaches the interface, the inner order parameter vanishes at the contact point, allowing the inner winding to change. Confinement by neighboring vortices can change which of the resulting states is thermodynamically favored.

A related axisymmetric Usadel calculation studied a centered vortex in a mesoscopic diffusive grain of radius $R_0$ with a finite-resistance boundary \cite{Golubov2026}. It found discontinuities in the pair potential, the supercurrent, and the local density of states. In the fully decoupled GL limit, a centered singly quantized vortex exists only for $R_0>1.84\,\xi_{\mathrm{GL}}(T)$, where $\xi_{\mathrm{GL}}(T)$ is the GL coherence length. This threshold concerns only the existence of the axisymmetric state, not its stability against translation or the conversion of an off-center core into an interfacial soliton.

Experimental work also motivates our study. Vortex spectroscopy can be used to estimate the transparency between grains \cite{Khapaev2025}. Scanning tunneling microscopy and spectroscopy resolve the spatially dependent vortex-core spectrum, as established by the pioneering measurements of Hess \textit{et al.} \cite{Hess1989STM,Hess1990STM} and by later thin-film studies \cite{Kashiwaya1992}. Discrete Usadel calculations for arrays of tunnel-coupled grains predict a quasiparticle minigap for a vortex centered between grains, when the coherence length is comparable to the grain spacing \cite{Kiselov2023}. Recent measurements on granular Nb report abrupt changes of the vortex-core structure at grain boundaries \cite{Stolyarov2026GranularNb}.

We therefore address three questions. First, can the closed interface capture a vortex from the surrounding superconductor? Second, is a captured centered core stable against translation? Third, can the interface absorb the point core and keep its winding in $\delta(\varphi)$? To answer them, we treat the inner and outer order parameters as independent fields.

We solve the two-dimensional GL problem near $T_c$ without restricting the form of the solutions. Appendix~A derives the finite-resistance GL interface conditions and the coefficient of the interface free energy from near-$T_c$ Usadel theory with Kupriyanov--Lukichev boundary conditions. We define $r_0=R_0/\xi_{\mathrm{GL}}(T)$ and the dimensionless interface resistance $\Gamma$. We then identify three main states: the centered vortex ($C$), the displaced vortex with a point core ($P$), and the interfacial relative-phase soliton ($S$).

The grain boundary attracts a distant vortex at any interface resistance (Sec.~\ref{subsec:isolated_map}), and the stability of the captured centered vortex depends on both the grain radius and $\Gamma$. We obtain two parameter-free results for an isolated vortex, within the symmetric GL model. First, in the nearly transparent limit ($\Gamma\to0$), the centered vortex is stable against small displacements for $r_0<r_0^*\simeq0.85$ (Appendix~B). Second, we denote by $\Gamma_{\mathrm{tr}}$ the interface resistance at which the centered vortex loses this local stability. For small grains, this threshold obeys $r_0\Gamma_{\mathrm{tr}}\to4/3$ as $r_0\to0$ (Appendix~C). Neither threshold follows from the phase-only models of closed weak links~\cite{Clem2010,Kaneyasu2010} or from the axisymmetric solution for a centered vortex~\cite{Golubov2026}. Both need the response of the core to a small shift, with the amplitude relaxing on both sides of the interface.

These results set the leading scale of the small-grain state diagram.
For an isolated grain, calculations of the displacement energy show
continuous $C/P$ onset followed by $P/S$ core absorption. We show that
the displaced core reaches the interface and is absorbed, and the unit
winding is then carried by $\delta(\varphi)$. Further, we show that in a
rigid one-vortex cell, a tricritical point separates continuous and
first-order vortex displacement. At larger grain radii the equilibrium
transition can occur directly from $C$ to $S$, with hysteresis. The
Supplemental Material~\cite{SupplementalMaterial} describes the numerical
methods, including the L-BFGS-B energy minimization~\cite{Byrd1995}.

\section{Model and isolated-vortex states}
\label{sec:model}

\begin{figure*}[t]
 \centering
 \includegraphics[width=0.9\textwidth]{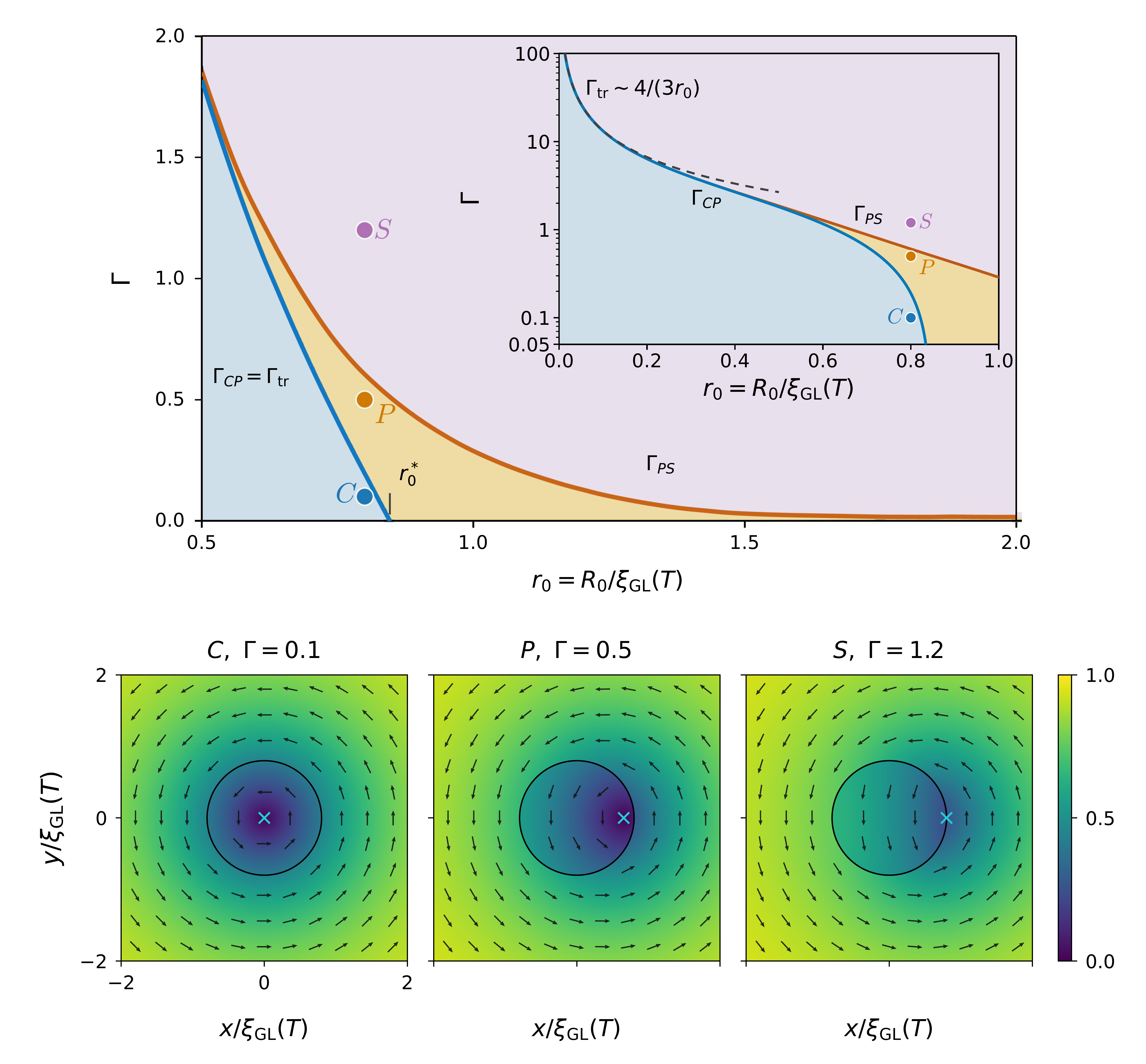}
 \caption{\textbf{Isolated-grain state diagram and representative vortex states.}
The upper panel shows the $C$, $P$, and $S$ states studied here as functions
of the dimensionless grain radius $r_0=R_0/\xi_{\rm GL}(T)$ and the
interface parameter $\Gamma$. Larger $\Gamma$ means weaker coupling
across the interface. 
The shaded regions apply for $\Gamma>0$. They do not include states with the
point core outside the grain, which can occur at very small $\Gamma$
(Sec.~\ref{sec:vortex_to_soliton}).
The blue curve is the isolated $C/P$ boundary and approaches $r_0^*\simeq0.85$ as $\Gamma\to0^+$.
Its label
$\Gamma_{CP}=\Gamma_{\rm tr}$ denotes continuous onset, where the
equilibrium boundary coincides with the centered state's translational
stability threshold. The line $\Gamma_{PS}$ marks absorption of
the displaced point core at the interface. The inset extends the diagram
to small $r_0$ on a logarithmic $\Gamma$ scale. The dashed curve shows
the leading small-grain asymptote $\Gamma_{\rm tr}\sim4/(3r_0)$
(Appendix~\ref{app:small_grain}). At
small $r_0$, the $P$ interval is too narrow to see on this scale.  The colored circles
at $r_0=0.8$ identify the states in the lower panel: $C$ at $\Gamma=0.1$,
$P$ at $\Gamma=0.5$, and $S$ at $\Gamma=1.2$. The lower panel shows
$|\Psi(x,y)|$ and the current directions in the window
$-2\le x/\xi_{\rm GL},y/\xi_{\rm GL}\le2$. Black circles mark the
interface. Cyan crosses mark the point zero in $C$ and $P$ and the center
of the relative-phase kink in $S$. The common color scale covers
$0\le|\Psi|\le1$ and is identical to the scale used for the
three-dimensional surfaces in the upper row of Fig.~\ref{fig:model_states}.}
 \label{fig:isolated_map}
\end{figure*}

\subsection{Ginzburg--Landau functional and interface conditions}
\label{subsec:model_geometry}

A circular grain of radius $R_0$ is embedded in a superconducting film of
identical material.  {The outer region contains no other
resistive interfaces.}  This symmetric choice isolates the effect of the
finite-resistance interface; real grain boundaries may also have different
bulk GL coefficients, conductivities, or local pair breaking on the two
sides.  A closed interface of finite resistance separates the inner and outer
regions.  Their pair potentials are independent fields, $\Psi_{\rm in}$ and
$\Psi_{\rm out}$.  One unit of winding is imposed on a remote contour, but the
vortex position is not fixed.

Lengths are measured in the GL coherence length,
$\mathbf r=\mathbf R/\xi_{\rm GL}(T)$.  The two domains are
\begin{equation}
 \Omega_{\rm in}=\{r<r_0\},\qquad
 \Omega_{\rm out}=\{r>r_0\},\qquad
 r_0=\frac{R_0}{\xi_{\rm GL}(T)}.
 \label{eq:model_domains}
\end{equation}
On the interface, $\varphi$ is the azimuthal angle, while $r_0^+$ and $r_0^-$ denote one-sided limits taken from the outer and inner regions, respectively.  The interface relative phase used throughout is
\begin{equation}
 \delta(\varphi)=
 \arg\Psi_{\rm out}(r_0^+,\varphi)
 -\arg\Psi_{\rm in}(r_0^-,\varphi).
 \label{eq:model_relative_phase}
\end{equation}
For the dirty-limit convention used here,
\begin{equation}
 \xi_s=\sqrt{\frac{\hbar D}{2\pi k_B T_c}},\qquad
 \xi_{\rm GL}(T)=\frac{\pi\xi_s}{2\sqrt{1-t}},\qquad
 t=\frac{T}{T_c},
 \label{eq:coherence_lengths}
\end{equation}
where $D$ is the normal-state diffusion coefficient.  We normalize the pair
potential by its uniform near-$T_c$ value,
\begin{align}
 \Psi_\mu&=\frac{\Delta_\mu}{\Delta_{\rm GL}(T)},
 \qquad \mu={\rm in,out},\nonumber\\
 \Delta_{\rm GL}^2(T)&=
 \frac{8\pi^2(k_B T_c)^2}{7\zeta(3)}(1-t).
\end{align}
Thus $|\Psi|=1$ in the uniform bulk state.

We use the dimensionless vector potential
$\mathbf A=(2\pi\xi_{\rm GL}/\Phi_0)\mathbf A_{\rm phys}$ and the covariant
derivative $\mathbf D=\boldsymbol\nabla-i\mathbf A$, where
$\Phi_0=h/(2e)$ is the superconducting flux quantum and
$\mathbf A_{\rm phys}$ is the physical vector potential.
{We neglect magnetic screening while retaining the full
amplitude and phase structure of the vortex.  In a bulk geometry the
relevant lateral scale $W$ must satisfy $W\ll\lambda$; with
$\kappa_{\rm GL}=\lambda/\xi_{\rm GL}$, this means
$\max(1,r_0)/\kappa_{\rm GL}\ll1$ near an isolated grain or
$r_s/\kappa_{\rm GL}\ll1$ in a finite cell of dimensionless radius $r_s$
(Sec.~\ref{sec:finite_cell_confinement}).  In a film with
$d\ll\lambda$, the corresponding condition is
$W\ll\Lambda=2\lambda^2/d$.  For the isolated calculation we set $\mathbf A=0$ and impose unit winding
at a remote contour.}
After subtraction of the uniform bulk contribution, the free energy is
\begin{align}
 F={}&\sum_{\mu={\rm in,out}}\int_{\Omega_\mu}
 \left[|\mathbf D\Psi_\mu|^2+
 \frac12\left(1-|\Psi_\mu|^2\right)^2\right]d^2r \nonumber\\
 &+\frac{1}{\Gamma}\oint_{r=r_0}
 |\Psi_{\rm out}-\Psi_{\rm in}|^2\,ds.
 \label{eq:model_functional}
\end{align}
For isolated unit-winding configurations,
$|\mathbf D\Psi|^2=r^{-2}+O(r^{-3})$ gives
$F_\alpha(\mathcal R)=2\pi\ln\mathcal R+C_\alpha+o(1)$, where
$\mathcal R$ is the remote cutoff and $C_\alpha$ the finite part for
configuration $\alpha$.  With identical remote boundary data, the logarithm
cancels from energy differences between configurations.
The dimensionless $F$ represents free energy per unit
vortex-line length in units of
\begin{equation}
 F_0(T)=N(0)\Delta_{\rm GL}^2(T)\xi_{\rm GL}^2(T)(1-t),
 \label{eq:energy_scale}
\end{equation}
where $N(0)$ is the normal-state density of states at the Fermi level
per spin and per unit volume.  Thus $F_0$ has units J\,m$^{-1}$ and the
energy in a film of thickness $d$ is $dF_0F$.
The dimensionless
interface resistance is
\begin{equation}
 \Gamma(T)=\frac{R_B}{\rho_s\xi_{\rm GL}(T)}.
 \label{eq:model_gammaB}
\end{equation}
Here $R_B$ is the specific resistance of the interface 
%--area product 
and $\rho_s$ is the normal-state
resistivity.  The transparent limit is $\Gamma\to0$.  Increasing $\Gamma$
weakens the coupling and permits a finite amplitude discontinuity and a nonzero relative phase $\delta(\varphi)$.  In terms of
the conventional Usadel parameter~\cite{KL},
\begin{equation}
 \gamma_B=\frac{R_B}{\rho_s\xi_s},\qquad
 \Gamma(T)=\frac{2\sqrt{1-t}}{\pi}\gamma_B.
\end{equation}
{Appendix A obtains Eq.~\eqref{eq:model_gammaB} and the
coefficient $1/\Gamma$ in Eq.~\eqref{eq:model_functional} from near-$T_c$
Usadel theory~\cite{Usadel1970} with Kupriyanov--Lukichev boundary
conditions~\cite{KL,GolubovKupriyanovIllichev2004}.}

Variation of Eq.~\eqref{eq:model_functional} gives
\begin{equation}
 -\mathbf D^2\Psi_\mu-\Psi_\mu+|\Psi_\mu|^2\Psi_\mu=0,
 \qquad \mu={\rm in,out},
 \label{eq:model_bulk_GL}
\end{equation}
and the interface conditions
\begin{align}
 \Gamma\,\mathbf n\!\cdot\!\mathbf D\Psi_{\rm in}
 &=\Psi_{\rm out}-\Psi_{\rm in},
 \label{eq:model_bc_in}\\
 \Gamma\,\mathbf n\!\cdot\!\mathbf D\Psi_{\rm out}
 &=\Psi_{\rm out}-\Psi_{\rm in}.
 \label{eq:model_bc_out}
\end{align}
The same geometrical normal $\mathbf n$ is used on both sides and points out
of the grain.  In the present dimensionless units, define
$\mathbf j_s^\mu=\operatorname{Im}(\Psi_\mu^*\mathbf D\Psi_\mu)$ and
$j_{s,n}^\mu=\mathbf n\cdot\mathbf j_s^\mu$ for $\mu={\rm in,out}$.  The
interface conditions conserve this supercurrent component  normal to the interface,
\begin{equation}
 j_{s,n}^{\rm in}=j_{s,n}^{\rm out}
 =\frac{1}{\Gamma}\operatorname{Im}
 \left(\Psi_{\rm in}^*\Psi_{\rm out}\right).
 \label{eq:model_current}
\end{equation}

\subsection{Topology and vortex states}
\label{subsec:model_states}

Three contours distinguish the relevant states,
\begin{align}
 N_{\rm in}&=\frac{1}{2\pi}\oint_{r=r_0-0}
 \boldsymbol\nabla\arg\Psi_{\rm in}\cdot d\mathbf l,\nonumber\\
 N_{\rm near}&=\frac{1}{2\pi}\oint_{r=r_0+0}
 \boldsymbol\nabla\arg\Psi_{\rm out}\cdot d\mathbf l,\nonumber\\
 N_\infty&=\frac{1}{2\pi}\oint_{C_\infty}
 \boldsymbol\nabla\arg\Psi_{\rm out}\cdot d\mathbf l=1.
 \label{eq:model_winding}
\end{align}
The contours do not pass through zeros of the order parameter.  Throughout
this work, $P$ denotes the displaced state with a point core. The representative states in the lower panel of
Fig.~\ref{fig:isolated_map} have
\begin{align}
 C &: (N_{\rm in},N_{\rm near},N_\infty)=(1,1,1),
 &&a=0,
 \label{eq:model_C}\\
 P &: (1,1,1),
 &&0<a<r_0,
 \label{eq:model_P}\\
 S &: (0,1,1),
 &&\text{no bulk point zero}.
 \label{eq:model_S}
\end{align}
Here $a$ is the core distance from the grain center. The conversion analyzed
below follows the interior-core route in Eq.~\eqref{eq:model_P}. Its distinction
from an exterior point-core state is discussed in
Sec.~\ref{sec:vortex_to_soliton}. In $S$, no bulk point zero remains in either
superconducting region, although the amplitude can be suppressed at the interface.

{For $P$, write $\mathbf a=a(\cos\varphi_a,\sin\varphi_a)$.  All
directions $\varphi_a$ are degenerate for an ideal circular interface.}
The displaced state therefore
breaks rotational symmetry and has a rotational zero mode.  Weak anisotropy or
disorder selects this direction.  The sharp centering bifurcation and its
quoted thresholds refer to the circular, identical-material model;
quantitative shifts caused by interface inhomogeneity are not calculated
here.

In $S$ the missing inner winding is carried by $\delta(\varphi)$ defined in Eq.~\eqref{eq:model_relative_phase}.  Its total variation is $2\pi$, as shown in Sec.~\ref{sec:vortex_to_soliton}.

\subsection{Centering and the $C$--$P$--$S$ diagram}
\label{subsec:isolated_map}

We distinguish the equilibrium $C/P$ boundary $\Gamma_{CP}$ from the
centered state's local translation threshold $\Gamma_{\rm tr}$.
The former is determined by equality of the competing minimum free
energies, $F_C=F_P$, whereas the latter is determined by a vanishing
quadratic energy cost of displacement. At a continuous $C/P$ onset the
two coincide.

Consider a small displacement $a$ of a centered isolated vortex.  We define
the interface-induced translational curvature $\lambda_{\rm int}$ and its
weak-interface coefficient $\chi(r_0)$ by
\begin{equation}
 F(a)-F(0)=\frac12\lambda_{\rm int}a^2+O(a^4),
 \qquad
 \chi(r_0)=\lim_{\Gamma\to0^+}
 \frac{\lambda_{\rm int}}{\Gamma}.
 \label{eq:model_chi}
\end{equation}
At $\Gamma=0$ the interface disappears, and translation invariance gives
$\lambda_{\rm int}=0$.  For a vortex displaced by $\mathbf a$, we define
\begin{align}
 I(a)&=\oint_{r=r_0}
 \left|\partial_n\Psi_v(\mathbf r-\mathbf a)\right|^2 ds,\nonumber\\
 \delta F_{\rm int}(a)&=-\Gamma I(a)+O(\Gamma^2),
 \label{eq:model_weak_interface}
\end{align}
where $\Psi_v=f(r)\exp(i\varphi)$ is the solution of the GL equation for an
unperturbed isolated vortex. 
\begingroup
For a nearly transparent interface, the coefficient of $a^2$ in the
free-energy difference $F(a)-F(0)$, calculated using the nonlinear Abrikosov
profile $f(r)$, changes sign once:
\par\endgroup

\begin{equation}
 r_0^*\simeq0.847,\qquad
 \chi(r_0)\begin{cases}
 >0,&r_0<r_0^*,\\
 <0,&r_0>r_0^*.
 \end{cases}
 \label{eq:model_pin_threshold}
\end{equation}
\begingroup
A small core displacement therefore raises the free energy for
$r_0<r_0^*$ and lowers it for $r_0>r_0^*$, making the centered vortex
locally stable below this radius and unstable above it.
\par\endgroup
  {The value $r_0^*$ is
fixed by the nonlinear radial GL vortex profile.  Appendix B gives the
weak-interface derivation and the numerical checks are described in the
Supplemental Material~\cite{SupplementalMaterial}.}

A remote exterior vortex is attracted to the grain. For
$a\gg\max\{r_0,1\}$, the linear phase response gives the interaction
energy relative to a vortex at infinity,
\begin{equation}
 U(a)\simeq-\frac{2\pi\Gamma r_0^2}{2r_0+\Gamma}\,\frac{1}{a^2}.
 \label{eq:model_exterior_attraction}
\end{equation}
In this far-field regime, the radial force is directed toward the grain
at arbitrary interface resistance. For $\Gamma\ll r_0$, Eq.~\eqref{eq:model_exterior_attraction}
reduces to $U(a)\simeq-\pi\Gamma r_0/a^2$; for $\Gamma\gg r_0$, its
coefficient saturates at $2\pi r_0^2$. This limit agrees with the far-field
pinning potential of an insulating cavity \cite{BespalovMelnikov2013}.
Supplemental Sec.~S2.2
\cite{SupplementalMaterial} gives the derivation.

We now ask when a vortex centered in an isolated small
grain (\(r_0\ll1\)) loses stability against a small displacement. At the
threshold \(\Gamma_{\rm tr}\), the quadratic displacement energy changes sign.
For a perturbation of fixed amplitude, the magnitude of the bulk energy
change inside the grain scales with its area,
\begin{equation}
 |\Delta F_{\rm bulk}|\sim r_0^2.
\end{equation}
A finite mismatch across the boundary costs interface energy proportional to
its circumference,
\begin{equation}
 \Delta F_{\rm int}\sim\frac{r_0}{\Gamma}.
\end{equation}
The two terms compete when
\begin{equation}
 \Gamma r_0=O(1).
 \label{eq:small_r_balance}
\end{equation}
This balance fixes the \(1/r_0\) scale. Appendix~C finds the coefficient
by matching the neutral displacement mode across the interface using two
decaying exterior solutions. The translation threshold is

\begin{equation}
 \lim_{r_0\to0}r_0\Gamma_{\rm tr}(r_0)=\frac{4}{3}.
 \label{eq:small_r_translation}
\end{equation}
The dashed curve in Fig.~\ref{fig:isolated_map} shows this leading
translation threshold, $\Gamma_{\rm tr}\sim4/(3r_0)$.
The limit is taken for an isolated vortex, with the remote boundary at
infinity, and with $g=r_0\Gamma$ fixed. Although $\Gamma\to\infty$,
the interface contribution remains comparable to the grain energy.
Decoupling at fixed $r_0$ is a different limit. A decoupled grain with a
centered vortex loses superconductivity for $r_0<1.84$, because the
supercurrent of the phase winding breaks pairs~\cite{Golubov2026}.

The upper panel of Fig.~\ref{fig:isolated_map} summarizes the isolated $C$, $P$, and $S$
states studied here. 
The blue $C/P$ boundary coincides with $\Gamma_{\rm tr}$ at
continuous onset and ends at $r_0^*$. The small-grain analysis gives a
positive relaxed quartic coefficient and a continuous $C/P$ onset in the
isolated limit. It predicts a narrow $P$ interval with
$\Gamma_{PS}-\Gamma_{\rm tr}\simeq1.48\,r_0^5$ to leading order.
Refined numerical continuation resolves this interval at the small radii
tested (Supplemental Sec.~S2.7~\cite{SupplementalMaterial}). The orange
curve tracks absorption of the displaced core at the interface. Unlike
$\Gamma_{CP}$, the line $\Gamma_{PS}$ is not a free-energy crossing:
it marks a change of winding numbers during core absorption, while the
fields and the free energy remain continuous (Sec.~\ref{subsec:core_interface}). The inset
compares the boundary curves with the leading translation asymptote. Competition with exterior point-core states at very small $\Gamma$ is
not included in the shading. Section~\ref{sec:vortex_to_soliton}
explains why this regime is beyond the scope of the present work.

\begin{figure*}[t]
 \centering
 \includegraphics[width=\textwidth]
 {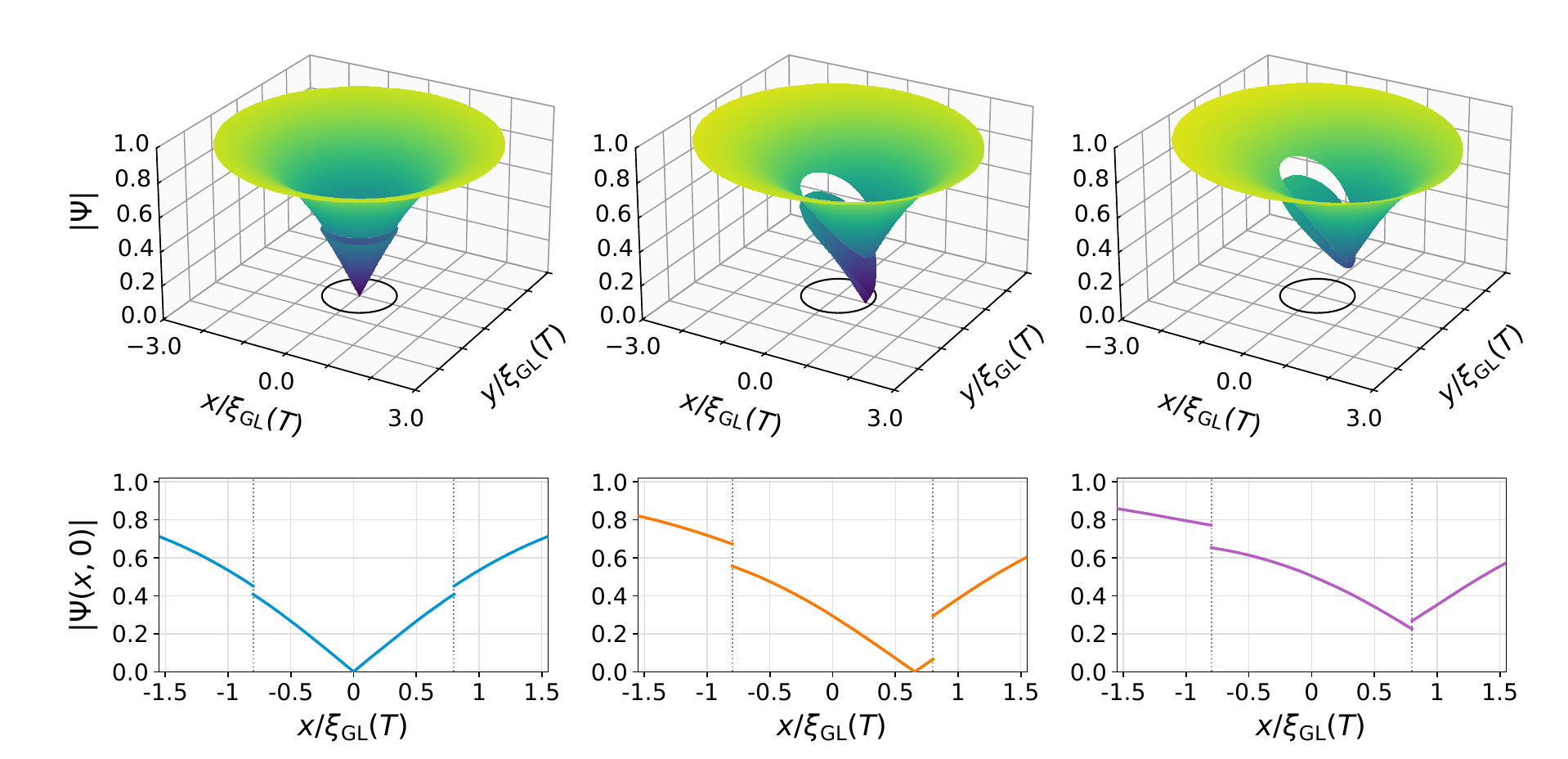}
 \caption{Representative order-parameter profiles corresponding to the colored circles in Fig. 1. From left to right, the columns show $C$ at $\Gamma = 0.1$, $P$ at $\Gamma = 0.5$, and $S$ at $\Gamma = 1.2$, respectively, all for $r_0 = 0.8$. Upper row: three-dimensional surfaces of the calculated amplitude $|\Psi(x,y)|$. The black circle marks the interface. The inner and outer surfaces are drawn separately to preserve the finite amplitude jump at the interface. The surfaces use the same scale $0 \le |\Psi| \le 1$ as the lower panel of Fig.~1. Lower row: line cuts $|\Psi(x,0)|$ along the reflection axis. Vertical dotted lines mark $x = \pm r_0$; the inner and outer one-sided curves are not joined across the finite-resistance interface. The displayed $C$ and $P$ states have $(N_{\mathrm{in}},N_{\mathrm{near}},N_{\infty})=(1,1,1)$, with the point core at the origin and at $r_{\mathrm{core}}\simeq 0.655$, respectively. The $S$ state has $(N_{\mathrm{in}},N_{\mathrm{near}},N_{\infty})=(0,1,1)$ and a finite inner-interface minimum $|\Psi|\simeq 0.225$.}
 \label{fig:model_states}
\end{figure*}

The lower panel compares all three states at the common radius
$r_0=0.8$, for which increasing $\Gamma$ passes through the $C$, $P$, and $S$
regions.  In $C$ the point zero is at the center and the current is rotationally
symmetric.  In $P$ the core moves off center while
$(N_{\rm in},N_{\rm near},N_\infty)=(1,1,1)$.  In $S$ no bulk point zero
remains; the strongest suppression and the largest azimuthal derivative of
the relative phase, $|\partial_\varphi\delta(\varphi)|$, are localized near
one point of the interface, where $\delta(\varphi)$ carries a $2\pi$ kink. 
The same
three solutions are shown as three-dimensional amplitude surfaces and
reflection-axis profiles in Fig.~\ref{fig:model_states}.

At $r_0=0.6$, the calculated $C/P$ onset is continuous. The translation
mode softens at $\Gamma_{\rm tr}=\Gamma_{CP}\simeq1.165$, with the
isolated-limit estimate and convergence checks given in the Supplemental
Material~\cite{SupplementalMaterial}. Denoting the translational curvature of the centered state by
$\lambda_{\rm tr}$ [defined as in Eq.~(\ref{eq:model_chi}); in the isolated
geometry $\lambda_{\rm tr}=\lambda_{\rm int}$], the near-onset behavior is
\begin{equation}
 \lambda_{\rm tr}\to0,\qquad
 a_P^2\propto\Gamma-\Gamma_{CP},\qquad
 F_C-F_P\propto(\Gamma-\Gamma_{CP})^2.
 \label{eq:isolated_scaling_main}
\end{equation}
{The translational curvature, displacement, and free-energy
scaling are shown in the Supplemental Material~\cite{SupplementalMaterial}.}

\section{Core-to-soliton conversion}
\label{sec:vortex_to_soliton}

\subsection{Displacement of the point core}
\label{subsec:translation}

At the continuous $C\to P$ onset, the core displacement starts from zero
and therefore initially lies inside the grain.  For the interior-core route
followed here, its position satisfies
\begin{equation}
 \Psi_{\rm in}(\mathbf r_v)=0,\qquad
 |\mathbf r_v|=a,\qquad 0<a<r_0.
 \label{eq:vortex_position}
\end{equation}
The displacement $a$ is measured from the converged field. It is not
imposed in the minimization. Increasing $\Gamma$ weakens phase locking
and lowers the cost of moving the core toward the boundary. Finite-cell
confinement changes the equilibrium $C/P$ onset and can make it first
order. In the calculated cells, a displaced point core is also absorbed
through a one-sided boundary zero.

\begin{figure}[t]
 \centering
 \includegraphics[width=\columnwidth]{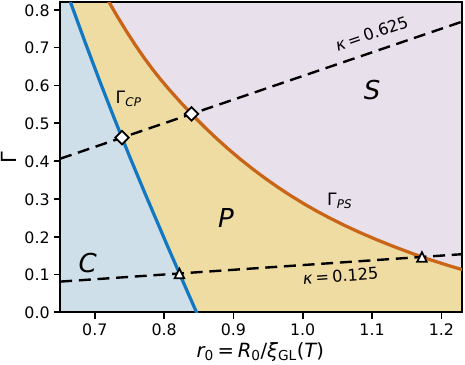}
 \caption{\textbf{Representative cooling paths through the isolated $C$--$P$--$S$
diagram.} The two dashed black lines are $\Gamma=\kappa r_0$ for
$\kappa=0.625$ and $0.125$. Diamonds and triangles mark the
intersections for $\kappa=0.625$ and $0.125$, respectively.}
 \label{fig:cooling_paths}
\end{figure}

At fixed physical $R_0$ and $R_B$, cooling below $T_c$ increases both
$r_0$ and $\Gamma$ in proportion to $\sqrt{1-T/T_c}$. Therefore, the trajectory is 
the straight path $\Gamma=\kappa r_0$, with
$\kappa=R_B/(\rho_sR_0)$. Figure~\ref{fig:cooling_paths} shows two representative cases.
For $\kappa=0.625$, both crossings lie near $r_0\simeq0.8$ and
$\Gamma\simeq0.5$, so the $P$ interval is narrow. For $\kappa=0.125$,
they span $r_0\simeq0.8$--$1.2$ and $\Gamma\simeq0.10$--$0.15$,
producing a substantially broader $P$ interval. Thus both paths give
continuous $C/P$ onset followed by $P/S$ core absorption for grains with
$R_0$ of order $\xi_{\rm GL}(T)$.

\begin{figure*}[!t]
 \centering
 \includegraphics[width=0.96\textwidth]
 {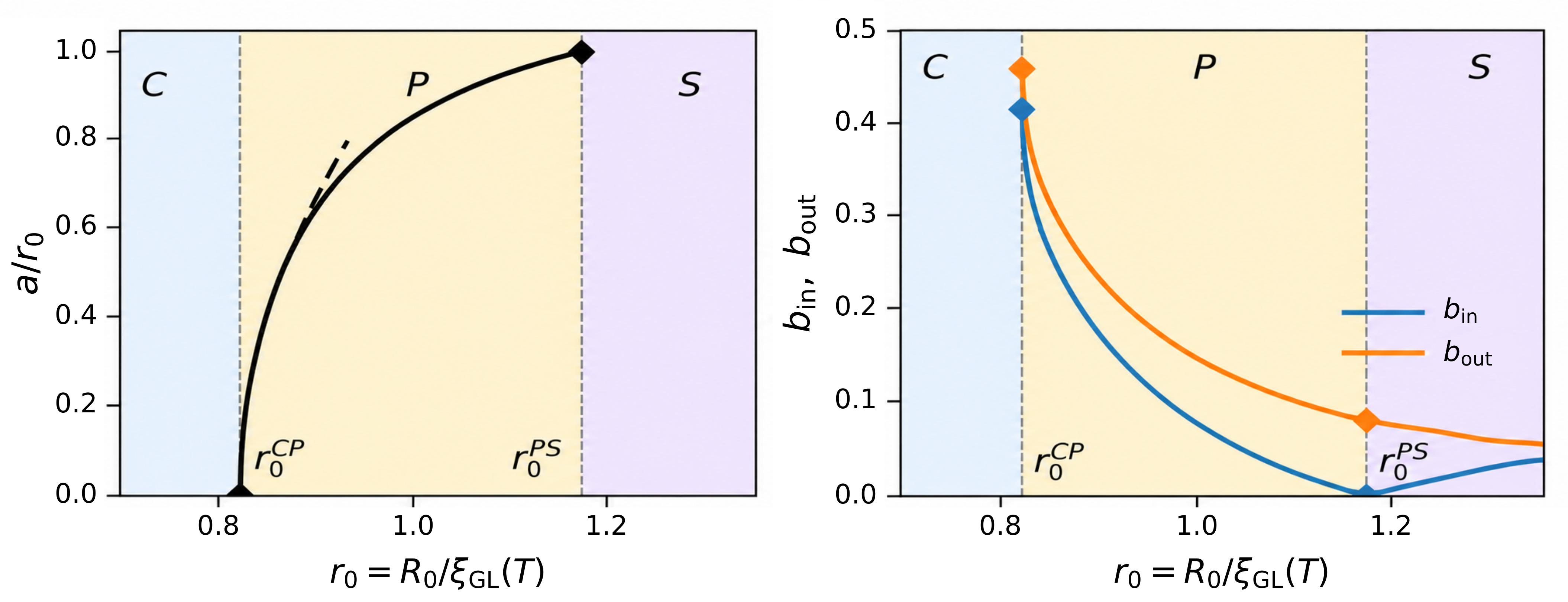}
 \caption{\textbf{Continuous core-to-soliton conversion along the isolated
 cooling path $\Gamma=0.125r_0$.}  Left panel: normalized radial position
 $a/r_0$ of the bulk point zero; the solid curve interpolates the calculated
 branch. The dashed curve shows the square-root guide
 $a/r_0=A\sqrt{r_0-r_0^{CP}}$ from $r_0^{CP}$ up to $a/r_0=0.8$.
 Right panel: one-sided interface amplitudes $b_{\rm in}$ and $b_{\rm out}$
 defined in Eq.~\eqref{eq:interface_amplitudes}. 
 Near $C/P$, the short segments continue the resolved numerical curves to the
 threshold with a square-root onset.
 At $P/S$,
 $a/r_0\to1$ and $b_{\rm in}\to0$, whereas $b_{\rm out}$ remains finite.
 The point-core coordinate ends at $a/r_0=1$ and is undefined in $S$.
 In the $S$ state, $b_{\rm in}$ rises continuously from zero while its minimum
 remains localized at the interface. Diamonds mark the $C/P$ and $P/S$
 boundaries; dashed vertical lines indicate
 $r_0^{CP}$ and $r_0^{PS}$.}
 \label{fig:path_resolved_conversion}
\end{figure*}

\subsection{Passage through the interface}
\label{subsec:core_interface}

As the displaced vortex evolves, its point core approaches the interface,
\begin{equation}
 a(\Gamma)\longrightarrow r_0.
 \label{eq:core_interface}
\end{equation}
At $\Gamma=\Gamma_{PS}$, the inner order parameter has a boundary zero at the
azimuth $\varphi_0$ where the core reaches the interface,
\begin{equation}
 \Psi_{\rm in}(r_0,\varphi_0)=0.
 \label{eq:interface_zero}
\end{equation}
The winding on a contour through this zero is not defined.  The inner winding
can therefore change without a singular event in either bulk region.  The
topology changes as
\begin{equation}
 (N_{\rm in},N_{\rm near},N_\infty):
 (1,1,1)\longrightarrow(0,1,1).
 \label{eq:conversion_winding}
\end{equation}
The point zero reaches the interface and is absorbed.  The imposed vorticity
is not lost; it is transferred to the interface relative phase
$\delta(\varphi)$.  To resolve the local modulus evolution, define
\begin{align}
 \varphi_{\min}&=\operatorname*{arg\,min}_{\varphi}
 |\Psi_{\rm in}(r_0^{-},\varphi)|,\nonumber\\
 b_{\rm in}&=|\Psi_{\rm in}(r_0^{-},\varphi_{\min})|,
 \qquad
 b_{\rm out}=|\Psi_{\rm out}(r_0^{+},\varphi_{\min})|.
 \label{eq:interface_amplitudes}
\end{align}
The outer amplitude is evaluated at the same angle $\varphi_{\min}$ as the
inner minimum.

Figure~\ref{fig:path_resolved_conversion} follows the physical cooling path
$\Gamma=0.125r_0$.  The direct path calculation, distinct from the guide
intersections in Fig.~\ref{fig:cooling_paths}, gives
\begin{align}
 (r_0^{CP},\Gamma_{CP})&=(0.821,0.103),\nonumber\\
 (r_0^{PS},\Gamma_{PS})&=(1.171,0.146).
 \label{eq:path_transition_points}
\end{align}
On the $P$ segment, $a/r_0$ increases continuously toward unity.  At the
$P/S$ point, the bulk zero reaches the inner side of the interface,
$a/r_0\to1$ and $b_{\rm in}\to0$, while
$b_{\rm out}=0.083$ remains finite.  Beyond this point the bulk point-core
coordinate is no longer defined, whereas $b_{\rm in}$ rises continuously from
zero in the $S$ state.  {On the numerical mesh used here, independently continued
$P$- and $S$-side solutions approach the same boundary-zero field and free
energy.  The mesh and continuity checks are given in Supplemental
Sec.~S2.6~\cite{SupplementalMaterial}.}

For the specific transition
$(N_{\rm in},N_{\rm near},N_\infty)=(1,1,1)\to(0,1,1)$, a nonvanishing
outer trace at the crossing point, together with the absence of outer zeros
elsewhere on the contour, is required to keep $N_{\rm near}=1$ well defined
while only $N_{\rm in}$ changes.  This is the topological role of the finite
$b_{\rm out}$.  It is not required to carry a nonzero normal current through
the boundary-zero point itself: Eq.~\eqref{eq:model_current} gives
$j_{s,n}(\varphi_{\min})=0$ exactly when $\Psi_{\rm in}=0$, even though
$\Psi_{\rm out}$ is finite.  Instead, the interface condition reduces locally
to $\Gamma\,\mathbf n\!\cdot\!\mathbf D\Psi_{\rm in}=\Psi_{\rm out}$,
so the finite outer amplitude produces a finite normal slope of the inner
field and permits a one-sided boundary zero.  Normal current can remain finite
at neighboring interface angles where both amplitudes are nonzero.
\subsection{Relative-phase soliton}
\label{subsec:soliton_physics}

After absorption, no bulk point zero remains in the grain.  The unit winding
at infinity is still present.  The mismatch is carried by the relative phase
in Eq.~\eqref{eq:model_relative_phase}, with
\begin{equation}
 \int_0^{2\pi}\partial_\varphi\delta(\varphi)\,d\varphi
 =2\pi\bigl(N_{\rm near}-N_{\rm in}\bigr)=2\pi.
 \label{eq:phase_soliton_winding}
\end{equation}
Thus the winding is stored as a $2\pi$ rotation of the relative
phase rather than in a bulk point core.

The $S$ state is therefore topologically equivalent to the Josephson-vortex
configuration of a closed (Corbino) weak link whose inner and outer
electrodes differ by one unit of phase winding~\cite{Clem2010,SanJose2026}:
the mismatch $N_{\mathrm{near}}-N_{\mathrm{in}}=1$ is carried by a $2\pi$ kink
of the phase difference along the junction. It differs from a Josephson
vortex in a long tunnel junction in three respects. Magnetic screening is
neglected here, so the kink carries the winding without a localized flux
quantum. The junction circumference $2\pi R_0$ is only a few
$\xi_{\mathrm{GL}}(T)$, so the kink extends over a sizable fraction of the
junction [see, e.g., the compact-cell profile in Supplemental Fig.~S8(b)]. Finally, the amplitudes on
both sides are obtained self-consistently and are suppressed at the kink
center, most strongly on the grain side [Fig.~\ref{fig:model_states}, right
column; compact cell: Supplemental Fig.~S8(a)].
The new element is therefore not the $S$ configuration itself, but its
continuous formation from an Abrikosov point core trapped inside the grain,
and the conditions under which this occurs.

The interface term itself separates exactly into amplitude mismatch and
relative-phase locking,
{\begin{align}
 \frac{|\Psi_{\rm out}-\Psi_{\rm in}|^2}{\Gamma}
 &=\frac{(|\Psi_{\rm out}|-|\Psi_{\rm in}|)^2}{\Gamma}\nonumber\\
 &\quad+J_{\rm eff}(s)[1-\cos\delta(\varphi)],\nonumber\\
 J_{\rm eff}(s)&=\frac{2|\Psi_{\rm in}||\Psi_{\rm out}|}{\Gamma}.
 \label{eq:interface_phase_locking}
\end{align}}
{Here $s=r_0\varphi$ is arclength along the grain boundary.}
Thus the local phase-locking energy vanishes at the one-sided boundary zero
because $b_{\rm in}=0$, even though $b_{\rm out}$ remains finite.  This is the
amplitude-assisted phase-slip event that changes $N_{\rm in}$.  {Away from the zero, interface locking favors $\delta=0$
modulo $2\pi$; the required $2\pi$ variation is localized in the
calculated $S$ state.}  A phenomenological local
phase-only illustration is
\begin{equation}
 F_\delta=\int ds\left[
 \frac{K_{\rm rel}}{2}\bigl(\partial_s\delta\bigr)^2
 +J\bigl(1-\cos\delta\bigr)\right].
 \label{eq:sine_gordon_effective}
\end{equation}
Here $K_{\rm rel}>0$ sets the energy cost of phase variations along the interface, while $J>0$ describes phase locking across it. This phase-only model is illustrative. In the full two-dimensional system, phase variations at different interface points are coupled through the superconductors on both sides \cite{Clem2010}. Changes in the magnitude of the order parameter are also essential when the point core disappears. We therefore obtain the interfacial soliton from the full GL functional, including both magnitude and phase variations. Its phase profile, for a compact-cell $S$ state, is shown in Supplemental Sec.~S5.5~\cite{SupplementalMaterial}.

A similar competition between an interfacial Josephson vortex and a
centered vortex was found for a cylindrical Ru inclusion in
Sr$_2$RuO$_4$~\cite{Kaneyasu2010}, using a phase-only interface model of
the same form as Eq.~\eqref{eq:sine_gordon_effective}. There, the phase
mismatch is imposed by the chiral order parameter of Sr$_2$RuO$_4$, the
interface coupling grows on cooling, and the interface vortex moves to the
center when the two energies cross. In our model, $\Gamma$ grows on
cooling, so the sequence runs in the opposite direction, from the centered
vortex to the soliton. Because the order-parameter modulus is not fixed
here, the GL calculation also resolves the displaced point-core state $P$ and
the continuous absorption of the core at the interface.

The calculated isolated $S$ state has winding numbers $(0,1,1)$ after
the inner point core is absorbed at the interface. An exterior point-core
state, $P_{\rm out}$, instead has $(0,0,1)$. Conversion between this
state and $S$ requires a zero of the outer field at the interface. This
is a different event from the inner-core absorption in Fig.~\ref{fig:path_resolved_conversion}. At the
representative $S$-state points sampled on both cooling trajectories,
the calculated energies favor $S$ over the exterior-core configurations
examined (Supplemental Sec.~S2.2~\cite{SupplementalMaterial}). 
At very small $\Gamma$, shallow exterior-core minima can also occur. The
energy differences between these minima and $S$ are too small to resolve
numerically. This regime is therefore beyond the scope of the present
work and needs a separate study.

Core absorption at $P/S$ proceeds continuously in these calculations.
Finite-cell confinement can instead select a direct first-order
equilibrium $C\to S$ crossing to an already-solitonic state
(Sec.~\ref{sec:phase_diagram} and Fig.~\ref{fig:phase_diagram}).

\section{Vortex transitions in a finite cell}
\label{sec:finite_cell_confinement}

\subsection{One-vortex cell}
\label{subsec:model_finite_confinement}

We now ask how confinement by neighboring vortices at finite magnetic
field affects the stability of a vortex at the grain center. We consider
a circular cell of radius $r_s$, centered on the grain and containing
one vortex. The superconductor outside the grain occupies the annulus
\begin{equation}
 \Omega_{\rm out}^{(r_s)}=\{r_0<r<r_s\},
 \label{eq:model_finite_domain}
\end{equation}
as sketched in Supplemental Sec.~S3~\cite{SupplementalMaterial}.
We choose the applied field so that one flux quantum passes through
the cell. In the symmetric gauge,
\begin{equation}
 \mathbf A(\mathbf r)=\frac{1}{r_s^2}(-y,x)
 =\frac{r}{r_s^2}\,\mathbf e_\varphi,
 \qquad
 \frac{H}{H_{c2}(T)}=\frac{2}{r_s^2}.
 \label{eq:model_vector_potential}
\end{equation}
At the cell edge, we fix the complex order parameter to its value
for a centered vortex,
\begin{equation}
 \Psi_{\rm out}(r_s,\varphi)=f_s e^{i\varphi}.
 \label{eq:model_outer_boundary}
\end{equation}
For each $(r_0,\Gamma,r_s)$, we obtain $f_s=f_{\rm out}(r_s)$
from the centered radial GL solution, using $f_{\rm in}(0)=0$,
the interface conditions, and $f'_{\rm out}(r_s)=0$.
The derivative condition applies only to this radial solution.
For a given $(r_0,\Gamma,r_s)$, we solve the two-dimensional GL
equations for each vortex state, keeping the order parameter fixed
at the same value $f_s e^{i\varphi}$ along the cell edge.
In this rigid-cell approximation, we compare the free energies of these solutions.

The phase imposed at the cell edge winds by $2\pi$, so every state
in the cell has the same total winding,
\begin{equation}
 N_{\rm cell}=\frac{1}{2\pi}\oint_{r=r_s}
 \boldsymbol{\nabla}\arg\Psi_{\rm out}\cdot d\mathbf l=1.
 \label{eq:model_cell_winding}
\end{equation}
This winding is preserved when the point core moves away from the
center and when it is absorbed at the grain interface. For the
centered vortex, the azimuthal superfluid velocity
$v_s=1/r-r/r_s^2$ vanishes at the cell edge.

For a small displacement of the centered vortex, we define the finite-cell
curvature $\lambda_{\rm tr}$ by the same free-energy expansion as in
Eq.~\eqref{eq:model_chi}. At fixed $r_0$ and $\Gamma$, it approaches
the isolated-grain curvature as the cell radius increases,
\begin{equation}
 \lim_{r_s\to\infty}\lambda_{\rm tr}(r_0,\Gamma;r_s)
 =\lambda_{\rm int}(r_0,\Gamma).
 \label{eq:model_confinement_decomposition}
\end{equation}
As shown in Sec.~\ref{subsec:phase_metastability}, the applied field and
the fixed cell boundary can stabilize the centered vortex against small
displacements ($\lambda_{\rm tr}>0$) even when it is unstable in the
isolated grain ($\lambda_{\rm int}<0$).

We use $r_s=16$ for the state diagram in
Sec.~\ref{sec:phase_diagram}. Supplemental Secs.~S5 and S7
discuss the smaller cell with $r_s=\sqrt{20}$~\cite{SupplementalMaterial}.

\subsection{Free energy and stability of the centered vortex}
\label{subsec:phase_regimes}
\label{subsec:phase_boundaries}

\begin{figure*}[t]
 \centering
 % v15 artwork: the left panel includes the green hatching and both marked crossings.
 \includegraphics[width=\textwidth]{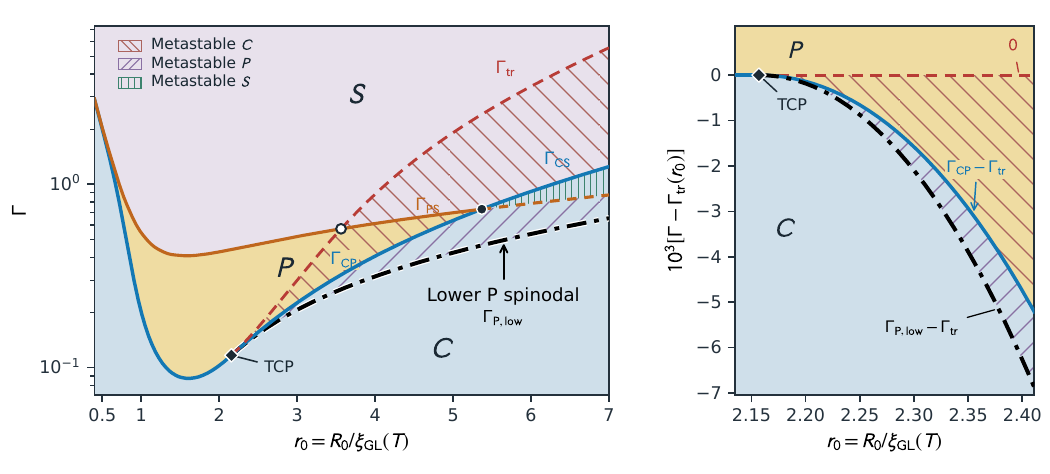}
\caption{\textbf{Equilibrium states and tricriticality in a rigid
 one-vortex cell.} The cell radius is $r_s=16$
 ($H/H_{c2}=2/r_s^2\simeq0.008$).
 Left panel: blue, yellow, and pink backgrounds show the equilibrium
 centered ($C$), displaced point-core ($P$), and interfacial-soliton ($S$)
 states. To the left of the tricritical point (TCP), the continuous $C/P$
 boundary is $\Gamma_{CP}=\Gamma_{\rm tr}$. To the right, the solid blue
 curve shows first-order $C/P$ or $C/S$ coexistence. The red dashed curve
 is the $C$ spinodal $\Gamma_{\rm tr}$. The black dash-dotted curve is
 the lower $P$ spinodal $\Gamma_{P,{\rm low}}$. The orange curve marks
 $P/S$ core absorption. The dashed part follows a metastable branch.
 The open circle at $(r_0,\Gamma)\simeq(3.56,0.57)$ marks
 $\Gamma_{\rm tr}=\Gamma_{PS}$. The filled circle at
 $(r_0,\Gamma)\simeq(5.37,0.73)$ marks the end of the equilibrium $P$
 region, where the blue and orange curves meet.
 Right panel: the boundaries near the TCP are shown on the scale
 $10^3[\Gamma-\Gamma_{\rm tr}(r_0)]$. The blue and black curves are
 $10^3(\Gamma_{CP}-\Gamma_{\rm tr})$ and
 $10^3(\Gamma_{P,{\rm low}}-\Gamma_{\rm tr})$, with all thresholds
 evaluated at the same $r_0$. The red dashed line is zero.
 Blue and yellow backgrounds show equilibrium $C$ and $P$; $S$ lies above
 this window. Red and violet hatching mark metastable $C$ and $P$ in
 both panels. Green vertical hatching in the left panel marks metastable
 $S$ alongside equilibrium $C$ for $\Gamma_{PS}<\Gamma<\Gamma_{CS}$.
 The two spinodals bound the hysteresis interval.}
  \label{fig:phase_diagram}
\end{figure*}

At fixed $r_0$, we increase $\Gamma$ across the $C/P$ boundary in
Fig.~\ref{fig:phase_diagram}. The equilibrium core displacement either
grows continuously from zero or jumps to a finite value. To distinguish
these cases, we fix the core at a displacement $a=|\mathbf a|$ and minimize
the GL free energy locally over the remaining degrees of freedom.
This gives $F_{\rm rel}(a)$, with $F_{\rm rel}(0)=F_C$.
By rotational symmetry, displacements of equal size in opposite directions
have the same energy. Near the center, the Landau expansion~\cite{LandauLifshitz1980}
gives
\begin{equation}
 F_{\rm rel}(a)-F_C=Aa^2+Ba^4+Da^6+O(a^8).
 \label{eq:phase_landau}
\end{equation}
As in Eq.~\eqref{eq:model_chi}, the translational curvature is the second
derivative of the free energy with respect to displacement. At the center,
$F_{\rm rel}''(0)=2A=\lambda_{\rm tr}$.

For $B>0$, the minimum moves continuously away from $a=0$ as $A$
changes from positive to negative [Fig.~\ref{fig:phase_landscapes}(a)].
The equilibrium transition then coincides with the loss of stability
of $C$, so $\Gamma_{CP}=\Gamma_{\rm tr}$. The continuous and first-order
parts of the $C/P$ boundary meet at the tricritical point. There
$A=B=0$ and $D>0$, so the leading energy cost is $Da^6$
[Fig.~\ref{fig:phase_landscapes}(b)].
For $B<0$ near this point, a displaced minimum can form while $C$
is still locally stable. In the sixth-order expansion, the two minima
have equal energy at $A=B^2/(4D)$, and the equilibrium displacement
jumps from zero to $\sqrt{-B/(2D)}$ [Fig.~\ref{fig:phase_landscapes}(c)].

A higher-energy local minimum represents a metastable state. The value
of $\Gamma$ at which it loses local stability is called a spinodal.
On increasing $\Gamma$ through a first-order $C/P$ transition, $C$
remains metastable for $\Gamma_{CP}<\Gamma<\Gamma_{\rm tr}$ and loses
stability at $A=0$. On decreasing $\Gamma$, $P$ loses stability at
$\Gamma_{P,{\rm low}}$, where the displaced minimum merges with the
saddle separating it from $C$. At this lower spinodal,
$\partial_a F_{\rm rel}=\partial_a^2 F_{\rm rel}=0$ at $a>0$.
In the sixth-order expansion this gives $A=B^2/(3D)$.
Both the displacement jump and the interval between the two spinodals
vanish as $B\to0^-$.

In the GL calculations, we locate first-order equilibrium transitions
by comparing the free energies at fixed grain radius $r_0$ and cell
radius $r_s$. At $C/P$ or $C/S$ coexistence, the two states share the
lowest free energy and satisfy, respectively,
\begin{align}
 F_P(r_0,\Gamma_{CP})&=F_C(r_0,\Gamma_{CP}),\nonumber\\
 F_S(r_0,\Gamma_{CS})&=F_C(r_0,\Gamma_{CS}).
 \label{eq:phase_energy_boundary}
\end{align}
We locate the centered stability limit from
\begin{equation}
 \lambda_{\rm tr}(r_0,\Gamma_{\rm tr};r_s)=0.
 \label{eq:phase_spinodal}
\end{equation}
The translation mode combines the $L=0$ and $L=2$ angular harmonics.
Supplemental Sec.~S7 gives the Landau coefficients and numerical
stability checks~\cite{SupplementalMaterial}. To locate
$\Gamma_{PS}$, we follow the point core in the full GL solution until
it reaches the grain interface.

\subsection{State diagram and hysteresis}
\label{sec:phase_diagram}
\label{subsec:phase_metastability}

Figure~\ref{fig:phase_diagram} shows the calculated state diagram for
$r_s=16$ ($H/H_{c2}\simeq0.008$). As $\Gamma$ increases, the sequence of
equilibrium states depends on the grain radius $r_0$. On the line
$\Gamma_{\rm tr}(r_0)$, $A=0$. We find the tricritical point where the
fitted coefficient $B$ changes sign and $D>0$. This gives
$(r_{0,\rm tri},\Gamma_{\rm tri})\simeq(2.16,0.12)$
(Supplemental Sec.~S7~\cite{SupplementalMaterial}).
For $r_0<r_{0,\rm tri}$, the core moves
continuously away from the center and is then absorbed at the interface.
The thresholds satisfy
\begin{equation}
 \Gamma_{CP}=\Gamma_{\rm tr}<\Gamma_{PS}.
 \label{eq:continuous_ordering}
\end{equation}
For $2.157<r_0<5.37$, the $C\to P$ transition is first order and is
followed by continuous $P/S$ core absorption. For $r_0>5.37$, the
equilibrium transition is directly from $C$ to $S$. The right panel of
Fig.~\ref{fig:phase_diagram} shows the equilibrium $C/P$ boundary
between the two spinodals on the first-order side of the tricritical point.

\begin{figure*}[t]
 \centering
 % v15 artwork: all six panels, including the configuration paths (d)--(f).
 \includegraphics[width=\textwidth]{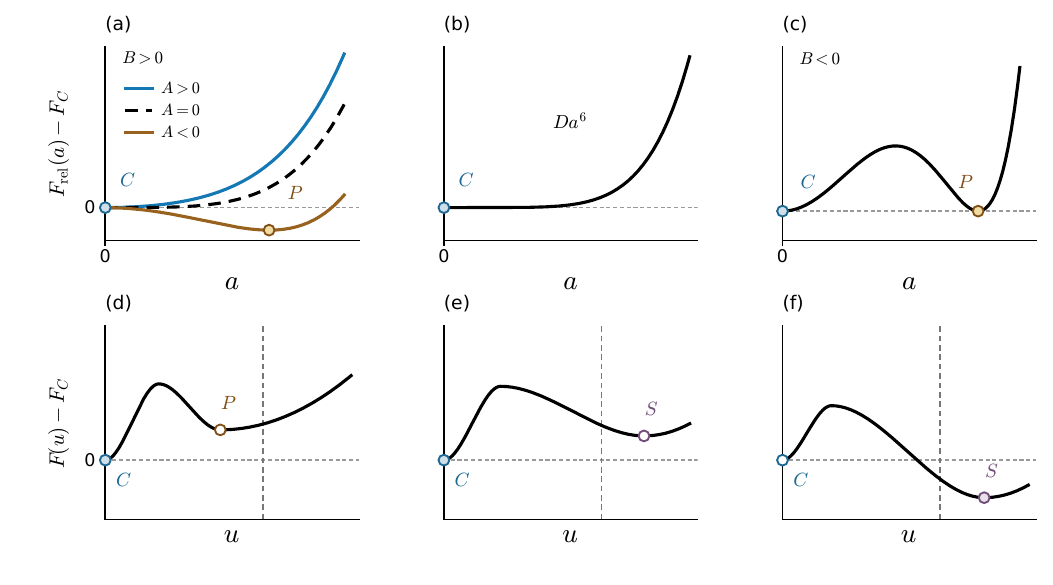}
 \caption{\textbf{Landau displacement profiles and schematic configuration
paths for Fig.~\ref{fig:phase_diagram}.} The top row illustrates the Landau expansion in
Eq.~\eqref{eq:phase_landau}: (a) Continuous $C/P$ onset for $B>0$ as $A$
passes through zero. (b) At the tricritical point, $A=B=0$ and $D>0$;
the leading term is $Da^6$. (c) First-order $C/P$ coexistence for $B<0$
at $A=B^2/(4D)$. Equal-depth $C$ and $P$ minima are separated by a
barrier. Bottom row: schematic paths beyond the end of the equilibrium
$P$ region in the left panel of Fig.~\ref{fig:phase_diagram}. (d) For
$\Gamma_{P,{\rm low}}<\Gamma<\Gamma_{PS}$, $C$ is the equilibrium state and
$P$ is metastable. (e) For $\Gamma_{PS}<\Gamma<\Gamma_{CS}$, $C$ is the
equilibrium state and $S$ is metastable. (f) For
$\Gamma_{CS}<\Gamma<\Gamma_{\rm tr}$, $S$ is the equilibrium state and
$C$ is metastable. Filled and open circles in the lower row mark
equilibrium and metastable minima. The coordinate $u$ labels
configurations at fixed $(r_0,\Gamma)$, not radial core positions. The
dashed vertical line marks inner-core contact ($a/r_0=1$ and $b_{\rm in}=0$).
To its right, $S$ configurations have $b_{\rm in}>0$ and no bulk point
core. Energy scales are schematic.}
 \label{fig:phase_landscapes}
\end{figure*}

Within the first-order $C/P$ regime, the centered stability limit
crosses the core-absorption line. Interpolating the calculated
thresholds gives $\Gamma_{\rm tr}=\Gamma_{PS}$ at
$(r_0,\Gamma)\simeq(3.56,0.572)$ (open circle in
Fig.~\ref{fig:phase_diagram}). On either side of this crossing,
\begin{equation}
 \begin{aligned}
 \Gamma_{P,{\rm low}}<\Gamma_{CP}<\Gamma_{\rm tr}<\Gamma_{PS}
       &\quad (2.157<r_0<3.56),\\
 \Gamma_{P,{\rm low}}<\Gamma_{CP}<\Gamma_{PS}<\Gamma_{\rm tr}
       &\quad (3.56<r_0<5.37).
 \end{aligned}
 \label{eq:first_order_CP_ordering}
\end{equation}
In both intervals, the equilibrium state changes from $C$ to $P$ and
then to $S$. In the first interval, $C$ loses local stability before
the core on the $P$ branch reaches the interface. In the second
interval, the core reaches the interface first.

We locate the end of the equilibrium $P$ region by requiring $F_P=F_C$
when the displaced core reaches the interface. This gives
$(r_0,\Gamma)\simeq(5.37,0.732)$ (filled circle in
Fig.~\ref{fig:phase_diagram}). At larger radii the thresholds are ordered as
\begin{equation}
 \Gamma_{P,{\rm low}}<\Gamma_{PS}<\Gamma_{CS}<\Gamma_{\rm tr}.
 \label{eq:first_order_CS_ordering}
\end{equation}
Panels (d)--(f) of Fig.~\ref{fig:phase_landscapes} show schematic
free-energy profiles at fixed $(r_0,\Gamma)$. Along each path, the
coordinate $u$ labels successive pairs of fields
$(\Psi_{\rm in},\Psi_{\rm out})$. We can therefore use $u$ even after
the point core disappears and its displacement $a$ is no longer defined.
As the core approaches the interface on the $P$ branch, $a/r_0\to1$
and $b_{\rm in}\to0$. In $S$, $b_{\rm in}>0$ and the relative phase
changes by $2\pi$ in a localized region of the interface. The same
solution branch continues from $P$ to $S$ whether it is in equilibrium
or metastable. At fixed $r_0$ and $r_s$, the calculated $F$ and
$\partial F/\partial\Gamma$ are continuous along this branch through
core absorption
(Supplemental Secs.~S5.4 and S6.1~\cite{SupplementalMaterial}).

For $r_0>5.37$ and $\Gamma_{P,{\rm low}}<\Gamma<\Gamma_{PS}$, $C$
is the equilibrium state and $P$ is metastable
[Fig.~\ref{fig:phase_landscapes}(d)].
At $\Gamma_{PS}$, the core is absorbed and the metastable $P$ state
becomes $S$. Its energy is still above $F_C$. This gives the green-hatched strip
$\Gamma_{PS}<\Gamma<\Gamma_{CS}$ in Fig.~\ref{fig:phase_diagram}
[Fig.~\ref{fig:phase_landscapes}(e)]. At $\Gamma_{CS}$, $C$ and $S$
have equal free energies. For $\Gamma_{CS}<\Gamma<\Gamma_{\rm tr}$,
$S$ is the equilibrium state and $C$ is metastable
[Fig.~\ref{fig:phase_landscapes}(f)].

Confinement changes the stability of the centered vortex. For example,
at $(r_0,\Gamma)=(0.8,0.5)$ the isolated vortex has a displaced core
[Fig.~\ref{fig:isolated_map}], while the centered state is stable
in the $r_s=16$ cell [Fig.~\ref{fig:phase_diagram}]. In the cell,
displacing the core deforms the surrounding condensate while the
order parameter at the cell edge stays fixed. The centered stability
limit $\Gamma_{\rm tr}$ first falls and then rises with $r_0$.
On the first-order side, the interval over which $C$ remains metastable
beyond the equilibrium transition has width
\begin{equation}
 \Delta\Gamma_C^{\rm meta}(r_0)=
 \begin{cases}
  \Gamma_{\rm tr}-\Gamma_{CP}, & C\leftrightarrow P,\\
  \Gamma_{\rm tr}-\Gamma_{CS}, & C\leftrightarrow S.
 \end{cases}
 \label{eq:phase_metastability_width}
\end{equation}
This interval widens at larger $r_0$. The cell size can also change the
equilibrium sequence. At $r_0=3$, reducing $r_s$ from $16$ to $\sqrt{20}$
removes the equilibrium $P$ interval and gives a direct $C\to S$ transition
(Supplemental Sec.~S7~\cite{SupplementalMaterial}).

The full hysteresis interval extends from $\Gamma_{P,{\rm low}}$ to
$\Gamma_{\rm tr}$ if the system stays in its local minimum until that
minimum loses stability. On an upward sweep with $2.157<r_0<3.56$,
$C$ jumps to $P$ at $\Gamma_{\rm tr}$. The core reaches the interface
later, at $\Gamma_{PS}$. When $\Gamma_{\rm tr}=\Gamma_{PS}$,
the core reaches the interface during the jump. For
$r_0>3.56$, $C$ jumps directly to $S$. Thus for $3.56<r_0<5.37$
an upward sweep can bypass the equilibrium $P$ state. On a downward
sweep from large $\Gamma$, $S$ converts continuously to $P$ at
$\Gamma_{PS}$, and $P$ can persist to $\Gamma_{P,{\rm low}}$, where
it jumps to $C$. Thermal activation can cause switching before a
spinodal is reached.

\section{Discussion and conclusions}
\label{sec:conclusions}
\subsection{Core reconstruction and confinement}
\label{subsec:discussion_mechanism}

In this work we have investigated how a closed resistive boundary around
a superconducting grain changes the structure and stability of a trapped
Abrikosov vortex. The calculations use two-dimensional Ginzburg--Landau
theory near $T_c$. A vortex can remain centered in the grain ($C$), move
off center while retaining its point core ($P$), or have its core absorbed
by the interface and become an interfacial relative-phase soliton ($S$).
In the last case, the point core disappears while the phase winding on
a contour outside the grain remains $2\pi$.

The results are organized in terms of
$r_0=R_0/\xi_{\rm GL}(T)$, the grain radius in coherence-length units,
and $\Gamma=R_B/[\rho_s\xi_{\rm GL}(T)]$, the dimensionless interface
resistance. Here $R_B$ is the specific interface resistance and $\rho_s$ is
the normal-state resistivity; larger $\Gamma$ means weaker coupling.
The diagrams compare the free energies of $C$, $P$, and $S$ and predict
 equilibrium transitions along any chosen path within the mapped
$(r_0,\Gamma)$ region. The two rays $\Gamma=\kappa r_0$ illustrate
$C/P/S$ conversion; other paths may give different transition sequences.

In the illustrated $C/P/S$ sequence, the displaced core reaches the
interface, where the inner order parameter vanishes at the contact
point. This allows its phase winding number to change from one to
zero. The outer order parameter remains nonzero around the interface
and keeps its unit phase winding. The difference between these winding
numbers is then carried by a $2\pi$ variation of the relative phase
around the interface. The fields and free energy change continuously
during core absorption (Fig.~\ref{fig:path_resolved_conversion}).
At representative $S$-state points, the energy comparisons also favor $S$
over the tested configurations with a point core outside the grain
(Supplemental Sec.~S2.2~\cite{SupplementalMaterial}). The regime of very small
$\Gamma$ is beyond the scope of this work and needs a separate study.

In the nearly transparent limit, $\Gamma\to0$,
an isolated centered vortex is locally stable against small displacements
for $r_0<r_0^*\simeq0.85$, and unstable above this radius (Appendix~B).
For very small grains, the resistance $\Gamma_{\rm tr}$ at which $C$
loses this local stability obeys $r_0\Gamma_{\rm tr}\to4/3$
(Appendix~C). The displacement then grows continuously, and core
absorption follows at a resistance higher by order $r_0^5$, so the
$P$ interval is very narrow (Supplemental Secs.~S2.5 and
S2.7~\cite{SupplementalMaterial}).

At finite field, we model confinement by neighboring vortices with
a circular cell containing the grain and one vortex. The applied field
is chosen to give one flux quantum through the cell, and the order
parameter is fixed at the outer boundary. The dimensionless cell radius is
$r_s=R_s/\xi_{\rm GL}(T)$. For $r_s=16$, the $C/P$ transition changes
from continuous to first order at a tricritical point. At still larger
grain radii, a displaced vortex can reach the interface and become a
soliton while its energy is still higher than that of $C$. These states
remain local energy minima and are therefore metastable. Core absorption
is continuous, but equilibrium stays centered until $S$ becomes favorable,
giving a direct first-order $C/S$ transition. The applied field and fixed
outer boundary can together stabilize $C$ where the isolated centered
vortex would be unstable, so
$r_0^*$ does not bound the $C$ region in the finite cell. Independent
finite-difference checks of representative compact-cell states support
this picture (Supplemental Secs.~S4 and S6~\cite{SupplementalMaterial}).

An energy barrier can keep $C$ metastable past the equilibrium transition,
allowing hysteresis. For a slow sweep that stays in the same local
minimum, the spinodals mark the limiting switching points, where that
minimum loses stability.
For example, the calculated $C/S$ transition path in the compact
$r_s=\sqrt{20}$ cell gives a dimensionless barrier
$\Delta F^\ddagger\simeq0.54$ at $r_0=3.5$, near coexistence at
$\Gamma_{CS}\simeq3.09$ (Supplemental
Sec.~S5.4~\cite{SupplementalMaterial}). For illustrative Nb parameters,
$T_c=7.3$~K, the material coherence scale $\xi_s=7$~nm, film thickness
$d=25$~nm, and density of states
$N(0)\simeq2\times10^{47}$~J$^{-1}$m$^{-3}$ per spin, the physical
barrier at $T=0.9T_c$ is about $20$~meV, or $35\,k_BT$
\cite{Stolyarov2026GranularNb,Kaplan1976}. To estimate a thermal
switching rate, we also need an attempt frequency. Disorder can change
both the barrier and the rate. The same energy unit, $dF_0\approx36$~meV
at $T=0.9T_c$, also sets the scale of thermal fluctuations,
$k_BT/(dF_0)\approx0.016$. Close to a continuous $C/P$ onset, where $F_C-F_P\ll k_BT/(dF_0)$, the
transition is thermally broadened: the core position fluctuates, and the
sharp mean-field thresholds become crossovers.

Near $T_c$,
$\xi_{\rm GL}(T)=\pi\xi_s/[2\sqrt{1-T/T_c}]$, so obtaining
$r_0=O(1)$ requires a grain sufficiently larger than $\xi_s$.
The barrier example above corresponds to $R_0\simeq120$~nm,
much larger than the grains in Fig.~\ref{fig:granular_gap_maps}.

\subsection{Relation to granular-film experiments}
\label{subsec:discussion_experiment}

Grain boundaries can pin vortices and modify their local
spectra~\cite{Volodin2002MFM,Hovhannisyan2025SQVM,Stolyarov2026GranularNb}.
The granular-Nb study of Ref.~\cite{Stolyarov2026GranularNb} measured
vortices at 1.1~K in a film with 3--10~nm grains. The reported coherence
scale, $\xi_{\rm exp}\simeq12$~nm, exceeds the grain size, so a vortex
spans several grains. These low-temperature observations motivate our
interface model, while the present GL calculation applies near $T_c$.

\begin{figure}[t]
 \centering
 \includegraphics[width=0.92\columnwidth]
 {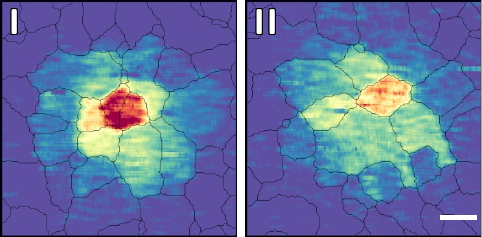}
 \caption{\textbf{Low-temperature excitation-gap maps of vortices in
 granular Nb.} Adapted from Ref.~\cite{Stolyarov2026GranularNb}:
 panels I and II show vortices ii and iii from the experimental column of
 its Fig.~1(c), at different sample locations. Colors show the STS gap
 $\Delta_{\rm STS}$ on the source scale 0--1.4~meV: red--orange
 indicates strong suppression, yellow--green intermediate values,
 and blue--violet values near the zero-field gap. Thin black lines mark
 grain boundaries, and the white scale bar is 10~nm. The maps alone
 do not determine whether the vortex is in the $C$, $P$, or $S$ state.}
 \label{fig:granular_gap_maps}
\end{figure}

Figure~\ref{fig:granular_gap_maps} shows two measured excitation-gap
patterns. One has strong gap suppression in its central grain; the
other retains a finite central gap and has a displaced low-gap region.
The excitation gap differs from the pair-potential amplitude $|\Psi|$
and does not directly reveal the order-parameter phase. The microscopic reconstruction
in Ref.~\cite{Stolyarov2026GranularNb} contains point phase singularities
even where the excitation gap is finite. The maps therefore cannot
identify $C$, $P$, or $S$ on their own.

STM studies of NbSe$_2$ found a bias-dependent apparent core radius
\cite{Volodin1997STM} and distinguished the spectroscopic core size
from the order-parameter core~\cite{Fente2016STM}. STS in Nb/Cu
bilayers also resolved enlarged cores in the normal layer
\cite{Stolyarov2018}.

For the Nb parameters above, the dirty-limit estimate used in
Ref.~\cite{Stolyarov2026GranularNb} gives
$\xi_s\simeq\xi_{\rm exp}\sqrt{\Delta_0/(6k_BT_c)}\simeq7$~nm,
with $\Delta_0\simeq1.35$~meV and $T_c=7.3$~K. Neither this scale
nor $\xi_{\rm exp}$ is the temperature-dependent $\xi_{\rm GL}(T)$
in Eq.~\eqref{eq:coherence_lengths}.

The model describes one closed resistive boundary in an otherwise
well-connected superconducting matrix. It shows how this boundary can
trap a vortex and absorb its point core while preserving the phase
winding outside the grain. The calculated transition boundaries and
metastability intervals may also depend on grain shape and on how the
surrounding condensate is represented, including the boundary conditions
at $r=r_s$. In a densely granular film, several
boundaries can affect the condensate within one coherence length.
Replacing their combined effect by uniform GL coefficients is an
additional approximation that we have not tested. The cell radius
$r_s$ represents confinement by neighboring vortices; it does not
account for the other grain boundaries. To compare with measurements
quantitatively, we need a model of the relevant grain network and a
microscopic calculation of its local quasiparticle density of states.

\subsection{Control parameters and experimental tests}
\label{subsec:discussion_tests}

An engineered circular weak link offers a direct test of the model:
its radius and coupling can be chosen independently of a random grain
network. Fabricated Corbino junctions demonstrate suitable geometries
and electrical control~\cite{Hadfield2003,Matsuo2020}. The link must
be narrow enough for the interface conditions to apply, and the
order-parameter amplitudes on both sides must be free to relax.

The parameter $\Gamma$ can be varied by changing the interface
resistance. Vortex spectroscopy can help estimate its transparency
\cite{Khapaev2025}, and experiments show that changing boundary
coupling can change the vortex arrangement~\cite{Wang2026GBEngineering}.
Temperature variation is one experimental way to trace a path through
this diagram. Near $T_c$, cooling reduces $\xi_{\rm GL}(T)$ and
therefore increases both $r_0$ and $\Gamma$. If $R_0$, $R_B$, and
$\rho_s$ are approximately constant, the path is a straight ray,
$\Gamma=\kappa r_0$, with $\kappa=R_B/(\rho_sR_0)$.
Following the same trapped vortex near $T_c$, at a low field where
vortices are well separated, could test the predicted conversion.
For example, $R_0=30$~nm, $\xi_s=7$~nm, and $\kappa=0.625$
[equivalently, $R_B/(\rho_s\xi_s)\simeq2.68$] give estimated
crossings at $T_{CP}/T_c\simeq0.93$ and $T_{PS}/T_c\simeq0.91$. These values
are illustrative. At fixed physical cell radius $R_s$, temperature
variation also changes $r_s$. Thus the $r_s=16$ diagram is a parameter
survey, not a temperature sweep of one fixed cell.

A phase-sensitive measurement could test
whether the inner winding changes to zero while the outer winding
stays equal to one. The calculated phase profile of a compact-cell $S$ state
is shown in Supplemental Sec.~S5.5 and Fig.~S8(b)~\cite{SupplementalMaterial}.
Spectroscopy can track the corresponding changes, while
scanning vortex microscopy and microwave microscopy can probe pinning
and the vortex response~\cite{Hovhannisyan2025SQVM,WangAnlage2025}.

Transport studies and time-dependent GL calculations for annealed
NbRe films show how grain boundaries guide vortex motion and the
growth of normal domains~\cite{MakhdoumiKakhaki2026NbRe}.
Coherent control and readout of vortex states in granular-Al
resonators~\cite{Nambisan2026VortexCoherence} also motivate studies
of vortex dynamics at resistive interfaces. These dynamical studies
do not identify the equilibrium $C$, $P$, or $S$ states considered here.

Our results show that a resistive grain boundary can change the structure
of a trapped Abrikosov vortex by absorbing its point core. The vortex
remains trapped, with its point core replaced by the Josephson-vortex
configuration of the closed grain boundary, a $2\pi$ kink of the interface
phase difference. Confinement by neighboring vortices can keep
the centered core metastable and allow a discontinuous, hysteretic
switch to the interfacial state. The interfacial state itself is the
known Josephson-vortex configuration of a closed weak link~\cite{Clem2010}. What
we add is how it forms: continuously, from a trapped Abrikosov core, with no
imposed mismatch. We also give the two parameter-free thresholds,
$r_0^*\simeq0.85$ and $r_0\Gamma_{\rm tr}\to4/3$, which control this process,
and the tricritical point and first-order switching caused by confinement. Such core
reconstruction provides a mechanism by which grain-boundary coupling can affect vortex pinning,
motion, and the associated dissipation in granular superconducting films.

\section*{Data availability}
% APS asks for a reason when data are not public, or a repository citation
% (e.g., Zenodo) if they are deposited. Copy this text into the submission form.
The data that support the findings of this article are not publicly
available. The data are available from the authors upon reasonable request.

\begin{acknowledgments}
This work was supported by the Russian Science Foundation, project no. 23-72-30004. The work presented in Section III was carried out as part of the state assignment of Lomonosov Moscow State University, as well as with the support of MSU Program of Development, Project
"Classical and quantum neuromorphic systems for high‑performance computing". 
\end{acknowledgments}

\appendix

% Appendix A: microscopic GL interface condition.
\begingroup
\section{Microscopic origin of the GL interface condition}
\label{app:usadel_gl_interface}

Here we connect the interface term in Eq.~\eqref{eq:model_functional} to
the diffusive Usadel--Kupriyanov--Lukichev theory
\cite{Usadel1970,KL,GolubovKupriyanovIllichev2004}.  The inner and outer
superconductors are the same material, with diffusion constant $D$ and
normal-state resistivity $\rho_s$; the boundary has specific resistance
$R_B$.  The unit normal $\mathbf n$ points from the grain to the outer
superconductor and is used on both sides of the boundary.

Write $\epsilon=1-T/T_c\ll1$, let
$\omega_n=(2n+1)\pi k_BT>0$, and define the physical-coordinate covariant
derivative
$\boldsymbol\Pi=\boldsymbol\nabla_{\!\mathbf R}
-i(2\pi/\Phi_0)\mathbf A_{\rm phys}$.
Near $T_c$ the anomalous Usadel Green function has the expansion
\begin{equation}
 f_n=\frac{\Delta}{\omega_n}
 +\frac{\hbar D}{2\omega_n^2}\boldsymbol\Pi^2\Delta
 -\frac{|\Delta|^2\Delta}{2\omega_n^3}
 +O(\epsilon^{5/2}).
 \label{eq:app_usadel_fexp}
\end{equation}
The gradient and cubic terms are both $O(\epsilon^{3/2})$.  Inserting
Eq.~\eqref{eq:app_usadel_fexp} into the self-consistency equation gives
\begin{equation}
 -\mathcal K\boldsymbol\Pi^2\Delta-\epsilon\Delta
 +\mathcal B|\Delta|^2\Delta=0.
 \label{eq:app_usadel_dimensional_gl}
\end{equation}
Here $\mathcal K=\pi\hbar D/(8k_BT_c)$ and
$\mathcal B=7\zeta(3)/[8\pi^2(k_BT_c)^2]$.
Thus $\xi_{\rm GL}^2=\mathcal K/\epsilon$ and
$\Delta_{\rm GL}^2=\epsilon/\mathcal B$, in agreement with the scales
defined in Sec.~\ref{subsec:model_geometry}.

To the same GL order, the anomalous component of the
Kupriyanov--Lukichev boundary condition reads
\begin{equation}
 \begin{aligned}
 \rho_s^{-1}\Pi_n f_{n,\rm in}
 &=\rho_s^{-1}\Pi_n f_{n,\rm out}\\
 &=\frac{f_{n,\rm out}-f_{n,\rm in}}{R_B}.
 \end{aligned}
 \label{eq:app_usadel_kl_f}
\end{equation}
Here $\Pi_n=\mathbf n\cdot\boldsymbol\Pi$.
At leading order $f_{n,\mu}=\Delta_\mu/\omega_n$; the common factor
$1/\omega_n$ cancels in \emph{each} of the two boundary conditions.  Defining
$\ell_B=R_B/\rho_s$ therefore gives the dimensional GL condition
\begin{equation}
 \ell_B\Pi_n\Delta_{\rm in}
 =\ell_B\Pi_n\Delta_{\rm out}
 =\Delta_{\rm out}-\Delta_{\rm in}.
 \label{eq:app_usadel_kl_delta}
\end{equation}
With $\Psi_\mu=\Delta_\mu/\Delta_{\rm GL}$ and
$\mathbf D=\xi_{\rm GL}\boldsymbol\Pi$, Eq.~\eqref{eq:app_usadel_kl_delta}
is precisely Eqs.~\eqref{eq:model_bc_in} and \eqref{eq:model_bc_out}, with
\begin{equation}
 \boxed{\Gamma(T)=\frac{R_B}{\rho_s\xi_{\rm GL}(T)}}.
 \label{eq:app_usadel_gamma}
\end{equation}

Equivalently, $\Gamma=\ell_B/\xi_{\rm GL}$ and
$\Gamma(T)=[2\sqrt{1-T/T_c}/\pi]\gamma_B$, where
$\gamma_B=R_B/(\rho_s\xi_s)$.
In particular, no additional Matsubara-sum coefficient multiplies
$\Gamma$.  The equal signs of the normal derivatives in
Eq.~\eqref{eq:app_usadel_kl_delta} follow from the \emph{same} geometrical
normal $\mathbf n$; the separate outward normals of the two GL domains
point in opposite directions.

%To fix the interface coefficient directly in the dimensionless functional
%\eqref{eq:model_functional}, let's write the interface term as
%$c\oint_{r=r_0}|\Psi_{\rm out}-\Psi_{\rm in}|^2\,ds$.
%Varying the two boundary values independently and including the surface
%terms from the bulk gradients gives
%$\mathbf n\cdot\mathbf D\Psi_\mu=c(\Psi_{\rm out}-\Psi_{\rm in})$
%on each side. Matching Eqs.~\eqref{eq:model_bc_in} and
%\eqref{eq:model_bc_out} fixes $c=1/\Gamma$.

The linear interface conditions are consistent with the nonlinear
bulk GL equation. The cubic bulk term enters the leading balance
because the quadratic bulk coefficient vanishes at $T_c$.
Nonlinear corrections to the interface conditions are of higher
order near $T_c$ and are omitted in the present GL approximation.

\endgroup

% Appendix B: weak-interface centering response.
% Entire new appendix marked blue for review; requires xcolor in the preamble.
\begingroup
\section{Weak-interface response and the centering radius}
\label{app:weak_interface}

In the isolated limit the remote boundary is at infinity. At
$\Gamma=0$ the interface disappears, and translating the vortex
leaves its energy unchanged. We calculate how a small positive
$\Gamma$ lifts this degeneracy. The unit-winding vortex at
$\Gamma=0$ has the profile
\begin{equation}
 \Psi_v=f(r)e^{i\varphi},\qquad
 f''+\frac{f'}r-\frac f{r^2}+f(1-f^2)=0,
 \label{eq:app_weak_profile}
\end{equation}
where $f(0)=0$ and $f(\infty)=1$.

Place the unperturbed vortex at displacement $\mathbf a$ from the grain
center. Let $\mathbf n$ point from the grain to the exterior, and set
$q=\partial_n\Psi_v(\mathbf r-\mathbf a)$ and
$\Delta\Psi=\Psi_{\rm out}-\Psi_{\rm in}$. The leading boundary
contribution to the change in the energy of Eq.~\eqref{eq:model_functional} is
\begin{align}
 &\oint ds\left(\frac{|\Delta\Psi|^2}{\Gamma}
       -2\operatorname{Re}[q^*\Delta\Psi]\right)\nonumber\\
 &\quad=\oint ds\left(\frac{|\Delta\Psi-\Gamma q|^2}{\Gamma}
       -\Gamma|q|^2\right).
 \label{eq:app_weak_square}
\end{align}
The term $|\Delta\Psi|^2/\Gamma$ is the interface energy. The linear term
comes from the first variation of the bulk energy, with the factor of two
from expanding the squared gradient. Since $\Psi_v$ solves the GL
equation in both domains, integration by parts leaves only boundary
terms. At the interface, the outward normals of the inner and outer
domains are $\mathbf n$ and $-\mathbf n$, respectively. The inner and
outer perturbations therefore enter with opposite signs, giving
$-2\operatorname{Re}[q^*\Delta\Psi]$.

Minimizing Eq.~\eqref{eq:app_weak_square} with respect to $\Delta\Psi$
gives $\Delta\Psi=\Gamma q+O(\Gamma^2)$ and the energy correction
\begin{equation}
 \delta F_{\rm int}(a)=-\Gamma I(a)+O(\Gamma^2),
 \label{eq:app_weak_energy}
\end{equation}
where $I(a)=\oint_{r=r_0}|\partial_n\Psi_v(\mathbf r-\mathbf a)|^2ds$.
%After the linear boundary term is included, the remaining contribution from relaxation of the bulk profile is of order $\Gamma^2$.

Equation~\eqref{eq:app_weak_energy} holds to first order in $\Gamma$ for
a vortex at any fixed displacement $a$ from the grain center.
To determine the stability of the centered vortex, we now expand $I(a)$
about $a=0$. 
When the vortex is centered, the radial phase gradient vanishes at the
interface. A displacement gives a radial phase gradient linear in $a$,
whose square contributes to $I(a)$ at the same order as the change in
the amplitude contribution. The phase contribution is
\begin{equation}
 \left.I(a)\right|_{\rm phase}=
 \frac{\pi a^2 f(r_0)^2}{r_0^3}+O(a^4).
 \label{eq:app_weak_gradient}
\end{equation}
Expanding the amplitude contribution and adding the phase contribution
gives (see Supplemental Sec.~S2.1~\cite{SupplementalMaterial})
\begin{align}
 I(a)-I(0)&=\frac{\pi r_0a^2}{2}K(r_0)+O(a^4),
 \label{eq:app_weak_I}\\
 K(r)&=G''+\frac{G'}r-\frac{2G}{r^2}+\frac{2f^2}{r^4}.
 \label{eq:app_weak_K}
\end{align}
Here $G(r)=f'(r)^2$. The radial GL equation~\eqref{eq:app_weak_profile}
gives an equivalent expression useful for determining the sign of $K$,
\begin{equation}
 K=2\left[(f'')^2+
   \left(\frac{f'}r-\frac f{r^2}\right)^2
   +f'^2(3f^2-1)\right].
 \label{eq:app_weak_K_rearranged}
\end{equation}
Combining Eqs.~\eqref{eq:app_weak_energy} and \eqref{eq:app_weak_I}
with the definition in Eq.~\eqref{eq:model_chi} gives the leading
curvature coefficient
\begin{equation}
 \chi(r_0)=\lim_{\Gamma\to0^+}
       \frac{\lambda_{\rm int}}{\Gamma}
       =-\pi r_0K(r_0).
 \label{eq:app_weak_chi}
\end{equation}
The regular core expansion gives $K(r)=-2f'(0)^2+O(r^2)<0$ near
the origin. Equation~\eqref{eq:app_weak_K_rearranged} gives $K>0$
wherever $f>1/\sqrt{3}$. Solving the radial GL equation numerically
(Supplemental Sec.~S2.1), we find a single zero,
\begin{equation}
 r_0^*\simeq0.847,\qquad K(r_0^*)=0.
 \label{eq:app_weak_radius}
\end{equation}
For fixed $r_0\ne r_0^*$ and sufficiently small $\Gamma>0$, the centered
vortex is stable against small displacements if $r_0<r_0^*$ ($\chi>0$)
and unstable if $r_0>r_0^*$ ($\chi<0$). At $r_0=r_0^*$,
the term linear in $\Gamma$ in the curvature vanishes. Thus
Eq.~\eqref{eq:app_weak_radius} gives the endpoint of the local
translational-stability boundary as $\Gamma\to0^+$. The calculated $C/P$ boundary
approaches the same endpoint. Its position at finite $\Gamma$ is
obtained by comparing the full two-dimensional free energies.

\endgroup

\section{Small-grain translation threshold}
\label{app:small_grain}
We consider an isolated vortex with winding \(+1\) at the center of
the grain. We take \(r_0\to0\) with
\(g=r_0\Gamma\) held fixed. At the threshold, a neutral displacement
along \(x\) takes the form
\begin{equation}
 v_\mu=u_{0,\mu}(r)+u_{2,\mu}(r)e^{2i\varphi},
 \qquad \mu=\mathrm{in,out},
 \label{eq:appC-sector}
\end{equation}
where \(u_{0,\mu}\) and \(u_{2,\mu}\) are real radial amplitudes.
The centered vortex has the profile
\(\Psi_{b,\mu}=f_\mu(r)e^{i\varphi}\). 
In the following radial equations, we omit the region index \(\mu\). The equations apply separately inside and outside the grain.
\begin{equation}
\begin{aligned}
 f''+\frac{f'}{r}-\frac{f}{r^2}+f(1-f^2)=0, \\
 u_0''+\frac{u_0'}{r}+(1-2f^2)u_0-f^2u_2&=0,\\
 u_2''+\frac{u_2'}{r}+
 \left(1-2f^2-\frac{4}{r^2}\right)u_2-f^2u_0&=0.
\end{aligned}
\label{eq:appC-neutral-system}
\end{equation}
The same interface condition applies to \(h=f,u_0,u_2\):
\begin{equation}
 \Gamma h_{\rm in}'=h_{\rm out}-h_{\rm in}
                   =\Gamma h_{\rm out}' ,
 \label{eq:appC-interface}
\end{equation}
Far from the grain, \(f_{\rm out}\to1\) and the displacement
perturbation vanishes.

\emph{Exterior solutions.---}
The exterior mode combines two solutions that vanish far from
the grain. The first is the translation
\(T=(f'+f/r,\,f'-f/r)^{\mathsf T}\) of the exterior profile
\(f=f_{\rm out}\) and decays as \(1/r\). The second solution,
\(S\), decays exponentially: far from the grain, \(S_0+S_2\)
behaves as \(K_0(\sqrt{2}\,r)\).
We therefore write
\begin{equation}
 \begin{pmatrix}u_0\\ u_2\end{pmatrix}=AT+BS.
 \label{eq:appC-two-responses}
\end{equation}
The regular vortex profile without the grain is denoted by \(f_{\rm v}\).
At fixed \(r>0\), \(f\) approaches \(f_{\rm v}\) as \(r_0\to0\).
Near the grain,
the radial equation gives terms proportional to \(r\) and \(1/r\).
Regularity inside the grain and the interface condition give
\(f'/f\simeq r_0/g\) at \(r_0\). Thus \(f\simeq c(r+r_0^2/r)\),
where \(c\) is the slope of \(f_{\rm v}\) at the origin.

Near the center of the limiting vortex, the \(L=2\) equation
reduces to Bessel's equation of order two. The singular part of
\(S_2\) behaves as \(Y_2(r)\propto r^{-2}\) and
its coefficient is absorbed into \(B\) in
Eq.~\eqref{eq:appC-two-responses}:
% This normalization requires a nonzero singular component in the
% independently decaying branch. If it vanished, the radial identity would
% also remove the logarithm. The resulting regular decaying mode
% would be proportional to T: setting w_L=S_L/T_L in the radial
% equations and integrating against w_L gives a sum of nonnegative
% terms (since T_0>0>T_2) that can vanish only for w_0=w_2 constant.
\begin{equation}
 S_2=\frac{1}{r^2}+O(1),\qquad
 S_0=k\ln r+O(1).
 \label{eq:appC-local-S}
\end{equation}
The two radial equations make $rW$ constant, where
$W=T_0S_0'+T_2S_2'-T_0'S_0-T_2'S_2$.
Both solutions decay at infinity, so this constant vanishes and
\begin{equation}
 T_0S_0'+T_2S_2'-T_0'S_0-T_2'S_2=0 ,
 \label{eq:appC-wronskian}
\end{equation}
For \(f_{\rm v}=cr-cr^3/8+O(r^5)\)
\cite[Sec.~5.1]{Tinkham1996},
\(T_0=2c+O(r^2)\) and \(T_2=-cr^2/4+O(r^4)\).
Equation~\eqref{eq:appC-wronskian} then gives
\(2ck+c/2+c/2=0\), hence \(k=-1/2\).

\emph{Matching at the interface.---}
The profile \(f\) is of order \(r\) down to the interface,
so matching to the actual exterior profile changes only
higher-order terms. Equations~\eqref{eq:appC-local-S} and
\eqref{eq:appC-wronskian} then give at \(r=r_0\)
\begin{equation}
 S_2\simeq r_0^{-2},\qquad S_2'\simeq-2r_0^{-3},
 \qquad S_0'\simeq-\frac{1}{2r_0},
 \label{eq:appC-matched-S}
\end{equation}
with \(S_0=O(|\ln r_0|)\). The symbol \(\simeq\) denotes
equality to leading order.

For the regular inner \(L=2\) mode,
Eq.~\eqref{eq:appC-interface} gives
\(u'_{2,\rm out}\simeq u_{2,\rm out}/(\Gamma+r_0/2)\)
at the interface. For \(\Gamma=g/r_0\), this net slope is small compared
with the separate exterior slopes.
Using \(T_2'\simeq2T_0/r_0\) and Eq.~\eqref{eq:appC-matched-S},
we obtain
\begin{equation}
 AT_2'\simeq -BS_2',
 \qquad B\simeq r_0^2 AT_0 .
 \label{eq:appC-l2-match}
\end{equation}

Let \(X\) and \(Y\) be the values of \(u_0\) just inside and
just outside the grain, respectively. At the interface,
Eq.~\eqref{eq:appC-two-responses} gives \(Y\simeq AT_0\), and
the equation for \(f\) gives \(T_0'=-f(1-f^2)\simeq-r_0T_0\).
The outer slope is then
\begin{equation}
 u'_{0,\rm out}\simeq -r_0Y-\frac{r_0Y}{2}
                    =-\frac{3r_0Y}{2}.
 \label{eq:appC-matched-slope}
\end{equation}
Inside the grain, \(f_{\rm in}=O(r_0^3)\), so the regular
\(J_0(r)\) solution gives \(u'_{0,\rm in}\simeq-r_0X/2\).
Matching the slopes gives \(X\simeq3Y\). Using this and
Eq.~\eqref{eq:appC-matched-slope} in the interface condition gives
\(2Y\simeq3gY/2\). Hence
\begin{equation}
 \lim_{r_0\to0}r_0\Gamma_{\rm tr}(r_0)=\frac43 .
 \label{eq:appC-small-grain-limit}
\end{equation}
Supplemental Sec.~S2.5 checks this limit and \(Y/X\to1/3\)
by solving the full radial equations numerically.

\bibliography{GL2D_main_revision_stage4}

\end{document}

\title{Supplemental Material for
``Vortex Core-to-Soliton Conversion in a Mesoscopic Superconducting Grain''}

\author{A.~A.~Golubov}
\affiliation{Moscow Institute of Physics and Technology, 141700 Dolgoprudny, Russia}
\affiliation{HSE University, 101000 Moscow, Russia}
\author{M.~Yu.~Kupriyanov}
\affiliation{Moscow Institute of Physics and Technology, 141700 Dolgoprudny, Russia}
\affiliation{Skobeltsyn Institute of Nuclear Physics, Lomonosov Moscow State University, Moscow 119991, Russian Federation}
\author{M.~M.~Khapaev}
\affiliation{Department of Numerical Methods, Faculty of Computational Mathematics and Cybernetics, Lomonosov Moscow State University, Moscow 119991, Russian Federation}
\author{A.~S.~Mel'nikov}
\affiliation{Moscow Institute of Physics and Technology, 141700 Dolgoprudny, Russia}
\affiliation{Institute for Physics of Microstructures, Russian Academy of Sciences, 603950 Nizhny Novgorod, GSP-105, Russia}
\author{V.~S.~Stolyarov}
\affiliation{Moscow Institute of Physics and Technology, 141700 Dolgoprudny, Russia}

%\date{8 November 2026}

\maketitle

\section*{Content and numerical methods}

\begingroup
We use the notation of the main text and measure lengths in
$\xi_{\rm GL}(T)$. The analytical centering and small-grain limits refer to
an isolated grain. The numerical isolated-grain boundaries use large
exterior domains. Finite-cell
calculations use $r_s=16$ for main-text Fig.~5 and $r_s=\sqrt{20}$ for the
strong-caging transition landscapes. Mesh parameters are given with each
calculation. 

The free-energy crossings $\Gamma_{CP}$ and $\Gamma_{CS}$, the centered-state
translation threshold $\Gamma_{\rm tr}$, and the core-contact (morphological)
boundary $\Gamma_{PS}$ describe distinct events. The analytical limit
$r_0\Gamma_{\rm tr}\to4/3$ is the dashed guide in main-text Fig.~1.
The small-grain calculation in Sec.~\ref{subsec:S_small_grain_onset}
supports $\Gamma_{CP}=\Gamma_{\rm tr}$ and predicts
$\Gamma_{PS}-\Gamma_{\rm tr}\sim1.48\,r_0^5$ in the isolated limit.
Thus $\Gamma_{CP}$, $\Gamma_{\rm tr}$, and $\Gamma_{PS}$ share the
leading coefficient $4/3$, although the $P$ interval has a nonzero
width at finite radius.

We find competing minima and transition paths by
finite-volume energy minimization in the compact cell. An independent
finite-difference solution of the GL boundary-value problem checks its
stationary fields and boundary ordering. For $r_s=16$ we use a radial
boundary-value calculation and polar variational continuation
(Sec.~\ref{sec:S_finite_size}).
\endgroup

% Microscopic origin of the GL equation and interface conditions.

\section{Usadel derivation and GL energy normalization}
\label{sec:S_usadel_gl}

Appendix A of the main text gives the near-$T_c$ reduction. Here we record
the intermediate Usadel step and the physical energy normalization. The two
regions are the same diffusive superconductor, with diffusion constant $D$,
normal-state resistivity $\rho_s$, and common $T_c$; their interface has specific
resistance $R_B$. The common
normal $\bm n$ points from the grain (in) to the surrounding film (out).

For $\omega_n=(2n+1)\pi k_BT>0$, the normalized Usadel Green function and the
gap obey~\cite{Usadel1970,GolubovKupriyanovIllichev2004}
\begin{align}
 \hbar D\,\widehat\nabla
    (\hat g_n\widehat\nabla\hat g_n)
 -[\omega_n\hat\tau_3+\hat\Delta,\hat g_n]&=0,
 \label{eq:S_usadel}\\
 \hat g_n^2&=\hat 1,\nonumber\\
 \Delta\ln\frac{T}{T_c}
 +2\pi k_BT\sum_{n=0}^{\infty}
 \left(\frac{\Delta}{\omega_n}-f_n\right)&=0.
 \label{eq:S_selfconsistency}
\end{align}
The physical-coordinate derivative acting on $f_n$ is
$\bm\Pi=\bm\nabla-i(2\pi/\Phi_0)\bm A_{\rm phys}$. With
\begin{equation}
 g_n=\frac{\omega_n}{\sqrt{\omega_n^2+\Phi_n\Phi_n^*}},
 \qquad f_n=\frac{g_n\Phi_n}{\omega_n},
 \label{eq:S_Phi_parameterization}
\end{equation}
normalization is automatic. The anomalous Usadel component and the identity
\begin{align*}
 \hbar D\,\bm\Pi\!\cdot\!
 (g_n\bm\Pi f_n-f_n\bm\nabla g_n)
 &=2(\omega_n f_n-g_n\Delta),\\
 g_n\bm\Pi f_n-f_n\bm\nabla g_n
 &=\frac{g_n^2}{\omega_n}\bm\Pi\Phi_n
\end{align*}
give the exact equation
\begin{equation}
 \Phi_n=\Delta+\frac{\hbar D}{2\omega_n g_n}
 \bm\Pi\!\cdot\!\left(g_n^2\bm\Pi\Phi_n\right).
 \label{eq:S_Phi_equation}
\end{equation}
For $\epsilon=1-T/T_c\ll1$, with
$\Delta=O(\epsilon^{1/2})$ and $\xi_s\Pi=O(\epsilon^{1/2})$, this yields
\begin{equation}
 f_n=\frac{\Delta}{\omega_n}
 +\frac{\hbar D}{2\omega_n^2}\Pi^2\Delta
 -\frac{\Delta|\Delta|^2}{2\omega_n^3}
 +O(\epsilon^{5/2}).
 \label{eq:S_fn_expansion}
\end{equation}
Substitution into the gap equation gives, to leading GL order,
\begin{equation}
 -\mathcal K\Pi^2\Delta-\epsilon\Delta
 +\mathcal B|\Delta|^2\Delta=0.
 \label{eq:S_dimensional_GL}
\end{equation}
Here $\mathcal K=\pi\hbar D/(8k_BT_c)$ and
$\mathcal B=7\zeta(3)/[8\pi^2(k_BT_c)^2]$.
Defining
\begin{equation}
 \Delta_{\rm GL}^2=\frac{\epsilon}{\mathcal B},
 \qquad \xi_{\rm GL}^2=\frac{\mathcal K}{\epsilon},
 \label{eq:S_GL_scales}
\end{equation}
and setting $\Psi=\Delta/\Delta_{\rm GL}$,
$\bm r=\bm R/\xi_{\rm GL}$, and $\bm D=\xi_{\rm GL}\bm\Pi$ gives
\begin{equation}
 -\bm D^2\Psi-\Psi+|\Psi|^2\Psi=0.
 \label{eq:S_dimensionless_GL}
\end{equation}

The anomalous Kupriyanov--Lukichev condition, linearized near $T_c$ with
the same normal on both sides, is~\cite{KupriyanovLukichev1988}
\begin{align}
 \rho_s^{-1}\Pi_n f_{n,\rm in}
 &=\rho_s^{-1}\Pi_n f_{n,\rm out}\nonumber\\
 &=\frac{f_{n,\rm out}-f_{n,\rm in}}{R_B},
 \qquad \Pi_n=\bm n\!\cdot\!\bm\Pi.
 \label{eq:S_KL_anomalous}
\end{align}
At leading order $f_{n,\mu}=\Delta_\mu/\omega_n$. The common factor
$1/\omega_n$ cancels, leaving the dimensionless conditions
\begin{align}
 \Gamma\,\bm n\!\cdot\!\bm D\Psi_{\rm in}
 &=\Psi_{\rm out}-\Psi_{\rm in},
 \label{eq:S_BC_in}\\
 \Gamma\,\bm n\!\cdot\!\bm D\Psi_{\rm out}
 &=\Psi_{\rm out}-\Psi_{\rm in}.
 \label{eq:S_BC_out}
\end{align}
Both signs are positive because the derivatives use one geometrical normal.
These conditions retain the leading near-$T_c$ order of the microscopic
interface condition.

The interface parameter and its relation to the conventional
Usadel parameter are
\begin{align}
 \Gamma(T)&=\frac{R_B}{\rho_s\xi_{\rm GL}(T)}
 =\frac{2\sqrt{\epsilon}}{\pi}\gamma_B,\nonumber\\
 \gamma_B&=\frac{R_B}{\rho_s\xi_s},\nonumber\\
 \xi_s&=\sqrt{\frac{\hbar D}{2\pi k_BT_c}}.
 \label{eq:S_Gamma_gammaB}
\end{align}

With $\bm j_s^\mu=\operatorname{Im}(\Psi_\mu^*\bm D\Psi_\mu)$ and
$j_{s,n}^\mu=\bm n\cdot\bm j_s^\mu$, the conditions also ensure
normal-current continuity:
\begin{align}
 j_{s,n}^{\rm in}=j_{s,n}^{\rm out}
 &=\frac{1}{\Gamma}
    \operatorname{Im}(\Psi_{\rm in}^*\Psi_{\rm out}).
 \label{eq:S_current_continuity}
\end{align}

\paragraph{Physical energy normalization.}
Let $N(0)$ be the normal-state density of states per spin and per unit
volume, so that $\rho_s^{-1}=2e^2N(0)D$. At the retained GL order, the
energy per unit vortex-line length, relative to the uniform superconducting
state, is
\begin{align}
\frac{F_{\rm phys}}{N(0)}
={}&\sum_{\mu}\int_{\Omega_\mu^{\rm phys}}\!d^2R\,
 \Bigl[\mathcal K|\bm\Pi\Delta_\mu|^2\nonumber\\
 &\hspace{15mm}
 +\frac{\mathcal B}{2}
   \bigl(|\Delta_\mu|^2-\Delta_{\rm GL}^2\bigr)^2
 \Bigr]\nonumber\\
 &+\frac{\mathcal K\rho_s}{R_B}\oint dS\,
 |\Delta_{\rm out}-\Delta_{\rm in}|^2.
 \label{eq:S_energy_physical}
\end{align}
Variation with respect to the independent fields yields
Eq.~\eqref{eq:S_dimensional_GL} and the common-normal interface
conditions. The energy unit and interface factor are
\begin{align}
 F_0&=N(0)\epsilon\Delta_{\rm GL}^2\xi_{\rm GL}^2,\nonumber\\
 \frac{\xi_{\rm GL}\rho_s}{R_B}&=\frac{1}{\Gamma}.
 \label{eq:S_energy_unit}
\end{align}
Then Eq.~\eqref{eq:S_energy_physical} reduces to the functional in main-text
Eq.~(5). Here $F_0$ has units J\,m$^{-1}$; a film of thickness $d$ has
energy $dF_0F$.

\section{Isolated interface: analytical limits and state diagram}
\label{sec:S_isolated}

We summarize the isolated-grain limits and the numerical construction of
the morphology diagram. The small-grain matching determines the centered
vortex's translation threshold; the equilibrium $C/P$ and core-contact
$P/S$ boundaries require separate calculations. In this section, $r_s$ is
the numerical outer cutoff used to approach the isolated limit, with
$\bm A=0$.
\subsection{Weak-interface centering threshold}
\label{subsec:S_response_kernel}

To evaluate the boundary integral $I(a)$ in Appendix B of the
main text, take $\mathbf a=a\hat{\mathbf x}$. On the circular
interface, the distance from the vortex core is
$\rho=(r_0^2+a^2-2ar_0\cos\varphi)^{1/2}$. Differentiating the
amplitude and phase in the radial direction gives
\begin{align}
 |\partial_r\Psi_v(\mathbf r-\mathbf a)|^2
 ={}&f'(\rho)^2\frac{(r_0-a\cos\varphi)^2}{\rho^2}\nonumber\\
 &+f(\rho)^2\frac{a^2\sin^2\varphi}{\rho^4}.
 \label{eq:S_weak_displaced_gradient}
\end{align}
For the angular integral, expand $\rho$ as
$r_0-a\cos\varphi+a^2\sin^2\varphi/(2r_0)+O(a^3)$ and use
$(r_0-a\cos\varphi)^2/\rho^2=1-a^2\sin^2\varphi/\rho^2$.
With $ds=r_0\,d\varphi$, this gives main-text Eqs.~(B4)--(B6).

To locate the zero in main-text Eq.~(B9),
the radial GL equation can be solved as a boundary-value problem on an adaptively refined radial mesh. We solve for $f/r$, which is
regular at the origin, using
\begin{gather}
 \left(\frac{f}{r}\right)''+
 \frac{3}{r}\left(\frac{f}{r}\right)'
 +(1-f^2)\frac{f}{r}=0,\nonumber\\
 \left.\left(\frac{f}{r}\right)'\right|_{r=0}=0,\nonumber\\
 f(R)=1-\frac{1}{2R^2}-\frac{9}{8R^4}-\frac{161}{16R^6}.
 \label{eq:S_vortex_profile_outer}
\end{gather}
The boundary value at the outer cutoff $R$ uses the large-$R$
expansion through order $R^{-6}$. We evaluate $K$ from the computed
profile using main-text Eq.~(B7). Changing the cutoff among
$R=25,35,45$ shifts the zero by less than $10^{-10}$.

% ---------------------------------------------------------------------
\subsection{Attraction of a remote exterior vortex}
\label{subsec:S_capture_classification}

We calculate the interaction energy $U(a)=F(a)-F(\infty)$ of a vortex with
winding $+1$ that lies in the film at distance $a$ from the grain center.
Here $F$ is the free energy of main-text Eq.~(5). All lengths are in units of
$\xi_{\rm GL}(T)$, magnetic screening is neglected, and
$a\gg\max(r_0,1)$. The result is main-text Eq.~(20).

Far from the vortex core, $|\Psi|=1$ and only the phase $\theta=\arg\Psi$
matters. Then main-text Eq.~(5) becomes
\begin{equation}
 F=\int|\nabla\theta|^2\,d^2r+\frac1\Gamma\oint\delta^2\,ds ,
 \label{eq:S_exterior_energy_functional}
\end{equation}
where $\delta=\theta_{\rm out}-\theta_{\rm in}$ at $r=r_0$ is the phase jump.
We used $|\Psi_{\rm out}-\Psi_{\rm in}|^2\simeq\delta^2$, which is valid
for small $\delta$.

Put the vortex at $\mathbf a=a\hat{\mathbf x}$ and use polar coordinates
$(r,\varphi)$ around the grain center. The phase of the vortex alone is
$\theta_v=\arg(z-a)$, with $z=x+iy$. Near the grain,
\begin{equation}
 \theta_v=\theta_0+G_v\,r\sin\varphi,\qquad G_v=-\frac1a ,
 \label{eq:S_exterior_incident_phase}
\end{equation}
with corrections of relative order $r_0/a$. Since $r\sin\varphi=y$, the vortex
produces a uniform phase gradient $G_v$ along $y$ across the grain. For
$|\Psi|=1$ the supercurrent density is proportional to $\nabla\theta$, so this is
also a uniform supercurrent through the grain.

For $|\Psi|=1$ the GL equation gives $\nabla^2\theta=0$ in both regions. The
two solutions with the angular dependence $\sin\varphi$ are $r\sin\varphi$ and
$\sin\varphi/r$. Inside the grain only the first is finite at the center.
Outside, the change of the vortex phase must vanish far from the grain, so only
the second is allowed. Therefore
\begin{equation}
 \theta_{\rm in}=\theta_0+c_{\rm in}\,r\sin\varphi,\qquad
 \theta_{\rm out}=\theta_v+\frac{d_{\rm dip}}{r}\sin\varphi .
 \label{eq:S_exterior_dipole_fields}
\end{equation}
Without the interface, $c_{\rm in}=G_v$ and $d_{\rm dip}=0$.

Varying Eq.~\eqref{eq:S_exterior_energy_functional} with respect to
$\theta_{\rm in}$ and $\theta_{\rm out}$ gives, at $r=r_0$,
\begin{equation}
 \Gamma\,\partial_r\theta_{\rm in}=\Gamma\,\partial_r\theta_{\rm out}=\delta .
 \label{eq:S_exterior_interface}
\end{equation}
These are the imaginary parts of main-text Eqs.~(10) and (11). The first
equality is the continuity of the supercurrent through the interface, because the
current is proportional to the phase gradient. The second equality says that this
current is proportional to the phase jump.
Equal gradients in Eq.~\eqref{eq:S_exterior_dipole_fields} give
$c_{\rm in}=G_v-d_{\rm dip}/r_0^2$, and the jump is
$\delta=(2d_{\rm dip}/r_0)\sin\varphi$. The condition
$\Gamma c_{\rm in}=2d_{\rm dip}/r_0$ then gives
\begin{equation}
 d_{\rm dip}=\frac{\Gamma G_v r_0^2}{2r_0+\Gamma},\qquad
 c_{\rm in}=\frac{2r_0}{2r_0+\Gamma}\,G_v .
 \label{eq:S_exterior_dipole_matching}
\end{equation}
The supercurrent is proportional to the phase gradient, so $c_{\rm in}/G_v$ is
the ratio of the supercurrent inside the grain to the supercurrent that the
vortex would drive there in a continuous film. The ratio is close to $1$ for
$\Gamma\ll r_0$, where the interface does not disturb the current, and close to
$0$ for $\Gamma\gg r_0$, where the current cannot cross the interface. The largest phase jump is
$2|d_{\rm dip}|/r_0\le 2r_0/a\ll1$, so the small-$\delta$ approximation holds
for any $\Gamma$.

To find the energy, write $\theta=\theta_v+\theta_1$. In a continuous film the
energy does not depend on $a$, so $U$ comes only from $\theta_1$:
\begin{equation}
 U=2\!\int\nabla\theta_v\!\cdot\!\nabla\theta_1\,d^2r+\int|\nabla\theta_1|^2d^2r
 +\frac1\Gamma\oint\delta^2ds .
 \label{eq:S_exterior_energy_split}
\end{equation}
By Eq.~\eqref{eq:S_exterior_dipole_fields},
$\theta_{1,\rm in}=-d_{\rm dip}\,y/r_0^2$ and $\theta_{1,\rm out}=d_{\rm dip}\,y/r^2$.

The first term is a boundary term. Since $\nabla^2\theta_v=0$ away from the
vortex, $\nabla\theta_v\!\cdot\!\nabla\theta_1=\nabla\!\cdot(\theta_1\nabla\theta_v)$,
and the divergence theorem turns each area integral into an integral over the
edge of its region. The edges are the interface, a small circle around the
vortex, and infinity. The last two give zero, because $\theta_v$ does not change
along the radius from the vortex and $\theta_1$ vanishes at infinity. At the
interface, the outward normal is $+\hat{\mathbf r}$ for the grain and
$-\hat{\mathbf r}$ for the exterior, and $\theta_{1,\rm out}-\theta_{1,\rm in}=\delta$
because $\theta_v$ is continuous. With $\partial_r\theta_v=G_v\sin\varphi$,
$\delta=(2d_{\rm dip}/r_0)\sin\varphi$, and $ds=r_0\,d\varphi$, we obtain
\begin{align}
 2\!\int\nabla\theta_v\!\cdot\!\nabla\theta_1\,d^2r
 &=2\oint(\theta_{1,\rm in}-\theta_{1,\rm out})\,\partial_r\theta_v\,ds
 \nonumber\\
 &=-2\oint\delta\,\partial_r\theta_v\,ds=-4\pi G_v\,d_{\rm dip}.
 \label{eq:S_exterior_cross}
\end{align}
This term is linear in $d_{\rm dip}$. It is the energy change caused by the
incident supercurrent crossing the phase jump. It is negative when $d_{\rm dip}$
has the same sign as $G_v$, so it favors the dipole, and the two quadratic terms
below limit its size.
Inside the grain, $|\nabla\theta_1|=d_{\rm dip}/r_0^2$ is constant, which gives
$\pi d_{\rm dip}^2/r_0^2$. Outside, $|\nabla\theta_1|^2=d_{\rm dip}^2/r^4$, which
gives $2\pi d_{\rm dip}^2\int_{r_0}^\infty r^{-3}dr=\pi d_{\rm dip}^2/r_0^2$.
The interface term is $4\pi d_{\rm dip}^2/(\Gamma r_0)$. The sum is
\begin{equation}
 U(d_{\rm dip})=-4\pi G_v\,d_{\rm dip}+\frac{2\pi d_{\rm dip}^2}{r_0^2}
 +\frac{4\pi d_{\rm dip}^2}{\Gamma r_0}.
 \label{eq:S_exterior_dipole_energy}
\end{equation}
Its minimum over $d_{\rm dip}$ is exactly the value in
Eq.~\eqref{eq:S_exterior_dipole_matching}, which checks the calculation. At the
minimum, the two quadratic terms equal half of the linear term with the
opposite sign, so $U=-2\pi G_v\,d_{\rm dip}$. With $G_v=-1/a$,
\begin{equation}
 U(a)\simeq-\frac{2\pi\Gamma r_0^2}{2r_0+\Gamma}\,\frac{1}{a^2},\qquad
 -\frac{dU}{da}\simeq-\frac{4\pi\Gamma r_0^2}{2r_0+\Gamma}\,\frac{1}{a^3}.
 \label{eq:S_exterior_attraction}
\end{equation}

The force points toward the grain for every $\Gamma>0$. For $\Gamma\ll r_0$,
$U\simeq-\pi\Gamma r_0/a^2$, which goes to zero as the interface becomes
transparent. For $\Gamma\gg r_0$, $U\simeq-2\pi r_0^2/a^2$. This is the leading
term of the exact result $2\pi\ln(1-r_0^2/a^2)$ for a vortex near an insulating
circular hole. The relative error of Eq.~\eqref{eq:S_exterior_attraction} is of
order $\max(r_0,1)^2/a^2$.

For representative parameter points in the shaded $S$ region of main-text
Fig.~1, we compared numerically the free energy of the interfacial soliton with
that of a vortex whose core is held outside the grain. The core was placed at
several positions, and the same grid ($r_s=192$) was used for both states. At every position tested, the soliton had the lower free energy, except at very
small $\Gamma$, where the numerical accuracy was not sufficient to resolve
the extremely small energy differences.
In this regime, shallow minima with the core just outside the grain can
therefore compete with $S$.
This regime is beyond the scope of the
present work and needs a separate study.

\subsection{Equilibrium $C/P$ boundary}
\label{subsec:S_CP_boundary}

The equilibrium $C/P$ boundary satisfies
\begin{equation}
 F_P(r_0,\Gamma_{CP})=F_C(r_0,\Gamma_{CP}).
 \label{eq:S_CP_definition}
\end{equation}
For a continuous transition, it coincides with the loss of centered-state
stability, $\Gamma_{CP}=\Gamma_{\rm tr}$.

\begin{figure*}[t]
 \centering
 \includegraphics[width=\textwidth]{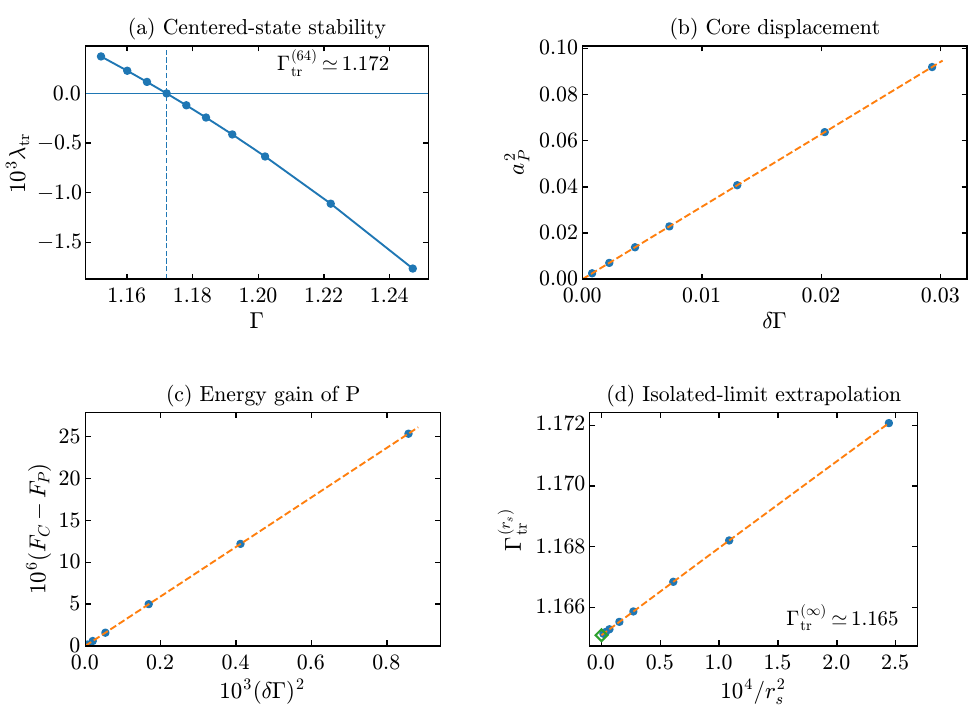}
 \caption{Continuous $C\to P$ transition at $r_0=0.6$.
 Panels (a)--(c) use $r_s=64$, with
 $\delta\Gamma=\Gamma-\Gamma_{\rm tr}^{(64)}$ and
 $\Gamma_{\rm tr}^{(64)}\simeq1.172$.
 (a) The translation stability parameter $\lambda_{\rm tr}$ of the
 centered state crosses zero.
 (b) The squared core displacement grows linearly with $\delta\Gamma$.
 (c) The energy gain $F_C-F_P$ grows quadratically with $\delta\Gamma$;
 plotting it against $(\delta\Gamma)^2$ gives a straight line.
 (d) Translation thresholds for $64\le r_s\le768$.
 Symbols show numerical results. Dashed lines in (b) and (c) show
 the near-onset fits; the dashed line in (d) is a quadratic fit in
 $r_s^{-2}$ using $r_s\ge96$. The open diamond marks the extrapolated
 isolated limit, $\Gamma_{\rm tr}\simeq1.165$.}
 \label{fig:S_isolated_cp_order}
\end{figure*}

Figure~\ref{fig:S_isolated_cp_order} illustrates this behavior at $r_0=0.6$.
For $r_s=64$, the translation stability parameter
$\lambda_{\rm tr}$ (Sec.~\ref{subsec:S_variational_continuation}) vanishes at
$\Gamma_{\rm tr}^{(64)}\simeq1.172$. Let $F_{\rm rel}(a)$ denote
the energy after relaxing the remaining order-parameter field at fixed
core displacement $a$. The numerical behavior near onset is described by
\begin{align}
 F_{\rm rel}(a)-F_C&=-\alpha_2\delta\Gamma\,a^2+\beta a^4+\cdots,
 \nonumber\\
 \delta\Gamma&=\Gamma-\Gamma_{\rm tr}^{(64)},
 \qquad \alpha_2>0,\quad\beta>0.
 \label{eq:S_CP_Landau}
\end{align}
Minimization with respect to $a$ gives
\begin{equation}
 a_P^2\propto\delta\Gamma,\qquad
 F_C-F_P\propto(\delta\Gamma)^2.
 \label{eq:S_CP_scaling}
\end{equation}
The calculated displacement and energy gain follow these laws:
the core moves continuously away from the center and $P$ has lower
energy. These results support
$\Gamma_{CP}^{(64)}=\Gamma_{\rm tr}^{(64)}$.

A quadratic fit in $r_s^{-2}$ to the translation thresholds for
$96\le r_s\le768$ gives the isolated-limit estimate
\begin{equation}
 \Gamma_{CP}(0.6)=\Gamma_{\rm tr}(0.6)\simeq1.165.
 \label{eq:S_CP_r0p6}
\end{equation}
The small-grain limit is treated separately in
Sec.~\ref{subsec:S_small_grain_onset}.

\subsection{Numerical $P/S$ boundary}
\label{subsec:S_PS_common_grid}

The orange $P/S$ line in main-text Fig.~1 is found by following the
displaced vortex until its core reaches the grain boundary. Choosing
the contact point at $\varphi=0$, we impose
\begin{equation}
 \Psi_{\rm in}(r_0,0)=0.
 \label{eq:S_inner_zero_constraint}
\end{equation}
Let $F_{\rm rel}(a;r_0,\Gamma)$ be the energy obtained by locally
relaxing the remaining field variables at fixed core position $a$.
At contact, the configuration must also be stationary with respect
to core displacement:
\begin{equation}
 Q(\Gamma)=
 \left.
 \frac{\partial F_{\rm rel}(a;r_0,\Gamma)}{\partial a}
 \right|_{a=r_0^-},
 \qquad Q(\Gamma_{PS})=0.
 \label{eq:S_contact_force}
\end{equation}
The derivative is taken from the $P$ side. Opposite signs of $Q$,
calculated while following the same locally relaxed solution,
bracket $\Gamma_{PS}$.

The mesh uses $r_s=192$ and $N_\varphi=256$. Radial spacing is
$h=0.04$ inside the grain and for 12 units outside it, then increases
to at most 1. The spatial discretization is described in
Sec.~\ref{subsec:S_variational_grid_energy}.
The gradient residual for the unconstrained field variables is below
$5\times10^{-8}$. Each $\Gamma$ bracket has relative width below
$1\%$ on this mesh. No spatial-grid extrapolation is applied, so this
bracket width is not a total error estimate.

\subsection{Numerical verification of the small-grain translation limit}
\label{sec:S25-local-matching}\label{subsec:S_small_r_extension}

At fixed $r_0$, we seek the resistance $\Gamma_{\rm tr}$ at which
the centered vortex has a neutral displacement mode. The bulk contribution
scales with grain area; the interface contribution scales with circumference
divided by $\Gamma$:
\begin{equation}
 \Delta F_{\rm grain}=O(r_0^2),\qquad
 \Delta F_{\rm int}=O(r_0/\Gamma).
 \label{eq:S25-energy-scales}
\end{equation}
The two terms compete when $g=r_0\Gamma=O(1)$.
This balance gives the scaling; Appendix C derives the coefficient
$4/3$. Here we specify the numerical problem and test that result.

We write the centered background as
$\Psi_{b,\mu}=f_\mu(r)e^{i\varphi}$, with
$\mu=\mathrm{in,out}$. In each domain it satisfies
\begin{equation}
 f_\mu''+\frac{f_\mu'}r-\frac{f_\mu}{r^2}
       +f_\mu(1-f_\mu^2)=0.
 \label{eq:S25-background}
\end{equation}
All lengths are in units of $\xi_{\rm GL}(T)$.
For a displacement along $x$, the perturbation contains the two
angular components
\begin{equation}
 \delta\Psi_\mu=\eta\bigl[u_{0,\mu}(r)
                    +u_{2,\mu}(r)e^{2i\varphi}\bigr],
 \label{eq:S25-mode-sector}
\end{equation}
where $u_{L,\mu}$ are real radial mode amplitudes.
At the neutral threshold their linearized equations are
\begin{equation}
 \begin{aligned}
 \left[-\partial_r^2-\frac1r\partial_r-1+2f_\mu^2\right]
             u_{0,\mu}+f_\mu^2u_{2,\mu}&=0,\\
 \left[-\partial_r^2-\frac1r\partial_r+\frac4{r^2}-1+2f_\mu^2\right]
             u_{2,\mu}+f_\mu^2u_{0,\mu}&=0.
 \end{aligned}
 \label{eq:S25-neutral-equations}
\end{equation}
The functions $u_L$ describe the displacement mode; $f_\mu$ is the
background vortex amplitude.

The background and both mode components obey the same interface law:
\begin{equation}
 \begin{aligned}
 \Gamma f'_{\rm in}&=f_{\rm out}-f_{\rm in}
                    =\Gamma f'_{\rm out},\\
 \Gamma u'_{L,\rm in}&=u_{L,\rm out}-u_{L,\rm in}
                    =\Gamma u'_{L,\rm out},\quad L=0,2.
 \end{aligned}
 \label{eq:S25-interface}
\end{equation}
Inner and outer values in this equation are evaluated at $r_0^-$
and $r_0^+$, respectively. Every prime denotes differentiation in
the increasing-$r$ direction, on both sides of the interface.
The traces are independent and may differ.

Regularity at the origin requires $f_{\rm in}=O(r)$,
$u'_{0,\rm in}(0)=0$, and $u_{2,\rm in}=O(r^2)$, with $u_{0,\rm in}$
finite. At infinity, $f_{\rm out}\to1$ and both mode components
decay. The exterior mode includes an algebraic translation tail,
\begin{equation}
 u_{0,\rm out}(r)=\frac{A}{r}+O(r^{-3}),\quad
 u_{2,\rm out}(r)=-\frac{A}{r}+O(r^{-3}).
 \label{eq:S25-translation-tail}
\end{equation}
Here $A$ is the coefficient of the translation solution $T$ in main-text
Eq.~(C4). It is fixed by interface matching and mode normalization.
At the finite numerical outer radius $R$, we retain the exact translation response
$T_0=f'_{\rm out}+f_{\rm out}/r$, $T_2=f'_{\rm out}-f_{\rm out}/r$,
and use the far-field decay $e^{-\sqrt{2}r}/\sqrt r$ for the second mode.
These give the two outer mode conditions.
Expanding Eq.~\eqref{eq:S25-background} in inverse powers of $r$ with
$f_{\rm out}\to1$ gives
$f_{\rm out}(R)=1-1/(2R^2)-9/(8R^4)-161/(16R^6)$.
Cutoffs $20\le R\le60$ were tested.

For each $r_0$, we vary $\Gamma$ and recompute the background until
Eqs.~\eqref{eq:S25-neutral-equations}--\eqref{eq:S25-interface}
admit a nonzero regular, decaying mode. This defines
$\Gamma_{\rm tr}(r_0)$. Neither $4/3$ nor the amplitude ratio $1/3$
is imposed. We measure
\begin{equation}
 \begin{aligned}
 g(r_0)&=r_0\Gamma_{\rm tr}(r_0),\\
 \rho(r_0)&=\frac{u_{0,\rm out}(r_0^+)}{u_{0,\rm in}(r_0^-)},\\
 \sigma(r_0)&=\frac{u'_{0,\rm out}(r_0^+)}
                   {r_0u_{0,\rm out}(r_0^+)}.
 \end{aligned}
 \label{eq:S25-numerical-definitions}
\end{equation}
Here $r_0^-$ and $r_0^+$ denote the inner and outer interface traces,
and the prime means differentiation in the increasing-$r$ direction.
The ratios $\rho,\sigma$ do not depend on the arbitrary mode scale.
The quantity $\rho$ measures the mode jump. The interface law gives
$\sigma=(1-\rho^{-1})/g$ exactly, so $\sigma$ is a derivative
consistency check rather than an independent constraint.
Appendix C predicts
\begin{equation}
 g\to4/3,\qquad \rho\to1/3,\qquad \sigma\to-3/2.
 \label{eq:S25-tr-asymptote}
\end{equation}

\begin{figure*}[t]
 \centering
 \includegraphics[width=0.9\textwidth]{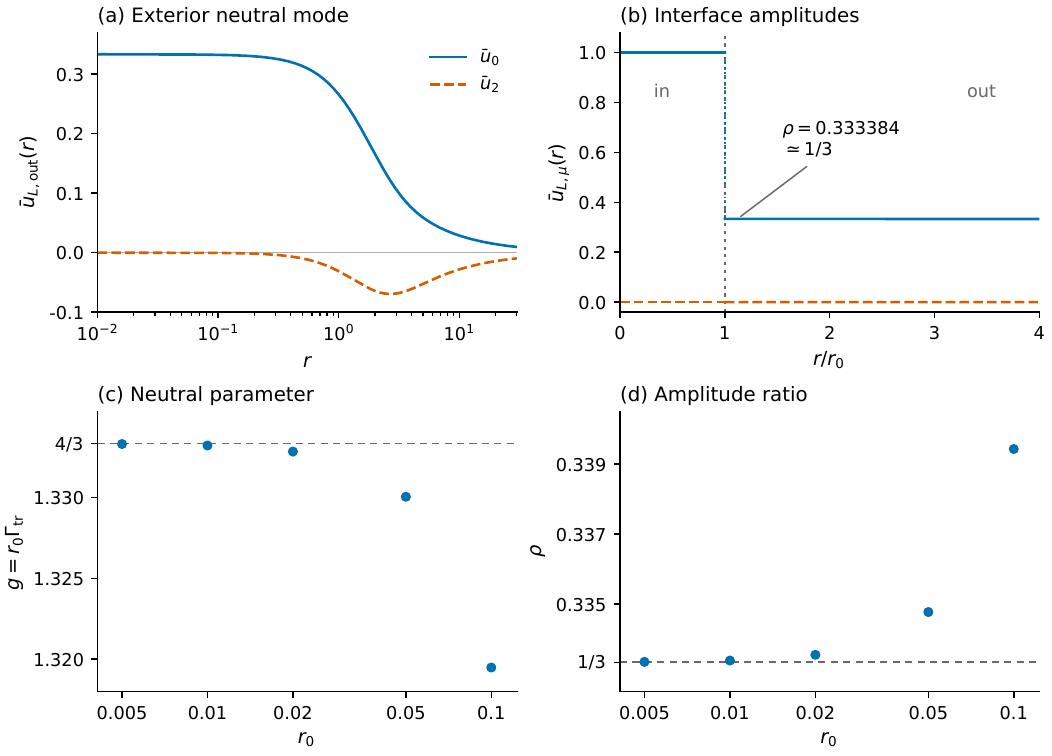}
 \caption{Self-consistent translation-neutral mode and small-grain
 convergence. The mode has
 $\delta\Psi_\mu=\eta[u_{0,\mu}+u_{2,\mu}e^{2i\varphi}]$;
 plotted amplitudes are
 $\bar u_{L,\mu}(r)=u_{L,\mu}(r)/u_{0,\rm in}(r_0^-)$.
 (a) Exterior $L=0,2$ functions at $r_0=0.01$ and
 $\Gamma_{\rm tr}\simeq 133.321$.
 (b) The same functions near the interface, located at $r/r_0=1$.
 The inner $L=0$ trace is one and the outer trace is
 $\rho=0.333384$, approaching $1/3$ for smaller grains.
 The small $L=2$ trace is nearly zero on this scale.
 (c) $g(r_0)=r_0\Gamma_{\rm tr}(r_0)$, the scaled neutral interface
 parameter.
 (d) $\rho(r_0)=u_{0,\rm out}(r_0^+)/u_{0,\rm in}(r_0^-)$,
 the outside-to-inside $L=0$ mode ratio.
 Dashed horizontal lines in (c),(d) mark $4/3$ and $1/3$.
 All radial coordinates are dimensionless.}
 \label{fig:S25-validation}
\end{figure*}

Figure~\ref{fig:S25-validation} shows the approach to the predicted small-grain limits.
Checks of numerical resolution, solver tolerance, and radial
cutoffs confirmed sufficient accuracy to resolve this convergence.

The approach of $\rho$ to $1/3$ is an outside-to-inside ratio of the
$L=0$ mode traces. The $L=2$ component is small near the interface but remains
part of the exterior displacement response.

\subsection{Representative cooling paths}
\label{subsec:S_cooling_rays}

At fixed $R_0$ and $R_B$, near-$T_c$ cooling follows $\Gamma=\kappa r_0$,
with $\kappa=R_B/(\rho_sR_0)$, neglecting weak temperature variations of
$R_B$ and $\rho_s$. The two example paths in main-text Fig.~3 show
continuous $C/P$ onset followed by $P/S$ core absorption. Reducing
$\kappa$ from $0.625$ to $0.125$ broadens the $P$ interval from
$\Delta r_0\simeq0.10$ to $0.35$.

Direct unrestricted continuation on $\Gamma=0.125r_0$, using
$r_s=192$, $h=0.04$, and $N_\varphi=256$, gives
\begin{align}
 (r_0^{CP},\Gamma^{CP})&\simeq(0.821,0.103),\nonumber\\
 (r_0^{PS},\Gamma^{PS})&\simeq(1.171,0.146).
 \label{eq:S_path_specific_points}
\end{align}

We extract the core coordinate $a$ when a bulk point zero is present and define
\begin{align}
 \varphi_{\min}&=\operatorname*{arg\,min}_{\varphi}
 |\Psi_{\rm in}(r_0^-,\varphi)|,\nonumber\\
 b_{\rm in}&=|\Psi_{\rm in}(r_0^-,\varphi_{\min})|,\nonumber\\
 b_{\rm out}&=|\Psi_{\rm out}(r_0^+,\varphi_{\min})|.
 \label{eq:S_path_amplitudes}
\end{align}
In main-text Fig.~4, the solid $a/r_0$ curve interpolates the calculated
branch and the dashed curve is the local law $a/r_0=A\sqrt{r_0-r_0^{CP}}$.
The amplitude curves use the matched onset near $C/P$ and resolved
continuation away from that immediate threshold region.

At $P/S$, $b_{\rm in}=0$ while $b_{\rm out}\simeq0.083$.
At offsets $r_0-r_0^{PS}=\pm2.5\times10^{-5}$, the inner amplitudes
are about $8\times10^{-6}$ and the outer trace remains smooth.
This check
supports continuous $P/S$ conversion on the computed branch.

\subsection{Small-grain displacement onset and width of the $P$ interval}
\label{subsec:S_small_grain_onset}

We first take $r_s\to\infty$ and then $r_0\to0$ with
$g=r_0\Gamma$ fixed. As in Appendix~C of the main text, $X$ denotes
the inner $L=0$ mode amplitude. It measures the nearly uniform
order-parameter perturbation inside the grain; $a$ is the core
displacement. Near onset, $X$ is proportional to $a$, and the leading
relaxed energy has the form
\[
 \frac{F_{\rm rel}(X)-F_C}{\pi r_0^2}
 =-\frac{g-g_{\rm tr}}{2}X^2+qX^4+\cdots,
\]
where $g_{\rm tr}=r_0\Gamma_{\rm tr}\to4/3$.
Numerical evaluation of the relaxed quartic coefficient gives
$q\simeq0.484>0$ in this limit.
Within the family of translated reference vortices used in this
expansion, no lower-energy state at finite displacement appears at
$g=4/3$.
These results predict a continuous $C/P$ onset with
$\Gamma_{CP}=\Gamma_{\rm tr}$ in the small-grain limit.

When the displaced vortex zero reaches the inner side of the interface,
the uniform component cancels the centered inner field at that point.
This gives $X_{\rm PS}\sim(3c/2)r_0^3$, where
$c=f_{\rm v}'(0)\simeq0.583$ is the central slope of the homogeneous
vortex profile. The outer-side order parameter remains nonzero.
Minimizing the energy above gives $g-g_{\rm tr}\sim4qX^2$ and hence
\[
 \Gamma_{\rm PS}-\Gamma_{\rm tr}
 \sim9c^2q\,r_0^5\simeq1.48\,r_0^5.
\]
Thus the $P$ interval has a positive but narrow width, although both
boundaries share the leading $4/(3r_0)$ behavior.

Numerical checks locate $\Gamma_{\rm tr}$ and $\Gamma_{\rm PS}$ on
the same grids and extrapolate their separation in $r_s^{-2}$.
At $r_0=0.10$, the isolated-limit width is $1.50\times10^{-5}$,
close to the asymptotic prediction. A relaxed $P$ state has lower
energy than $C$ within this interval, and its onset agrees with the
translation threshold. These results support a narrow equilibrium
$P$ interval in the resolved cases.

\FloatBarrier

\section{Numerical minimization and transition criteria}
\label{sec:S_variational}

\begin{figure}[t]
 \centering
 \includegraphics[width=0.52\linewidth]{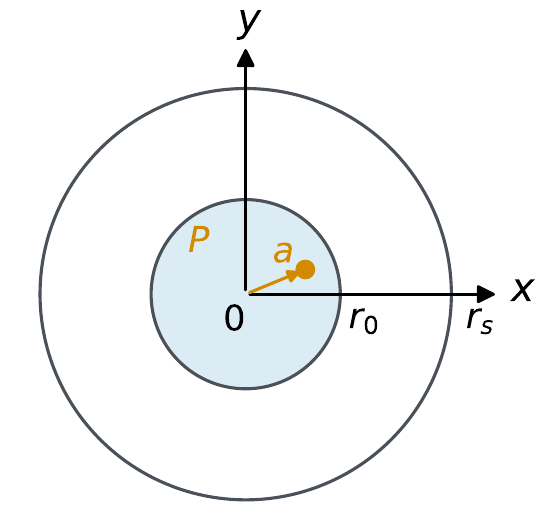}
 \caption{Circular cell containing one vortex. The grain and cell radii are
 $r_0$ and $r_s$; the orange point shows a displaced core at distance $a$
 from the center. Its direction is arbitrary. One flux quantum through
 the cell gives $H/H_{c2}=2/r_s^2$.}
 \label{fig:S_finite_cell_geometry}
\end{figure}

We minimize the GL free energy to find competing vortex states, compare
their energies, and calculate transition barriers.
Figure~\ref{fig:S_finite_cell_geometry} defines the cell geometry and
core displacement. The finite-cell calculations in
Secs.~\ref{sec:S_variational}--\ref{sec:S_additional_diagnostics} use
$r_s=\sqrt{20}$. Section~\ref{sec:S_finite_size} describes the
calculations at $r_s=16$.

\subsection{Spatial grid and boundary conditions}
\label{subsec:S_variational_grid_energy}

The grain and surrounding annulus have separate polar grids, with
$N_{r,{\rm in}}$ and $N_{r,{\rm out}}$ radial intervals and $N_\varphi$
angular points. The two grids meet exactly at the interface. The field values
on its two sides are independent and coupled only through the interface
energy. At the origin, one complex field value represents the single
physical point; it is allowed to vary, so a displaced vortex is not forced
to retain a zero at the center.

At the outer boundary we fix
\begin{equation}
 \Psi_{{\rm out},j}(r_s)=f_s e^{i\varphi_j},
 \qquad \varphi_j=\frac{2\pi j}{N_\varphi},
 \label{eq:S_var_outer_boundary}
\end{equation}
where $f_s$ comes from the centered radial solution at the same
$(r_0,\Gamma,r_s)$, obtained with $f'_{\rm out}(r_s)=0$ (main-text
Sec.~IV\,A). All states compared at these parameters use this same
boundary field, while the interior field is free to relax.

We evaluate the bulk GL energy by a finite-volume method. Gradients are
approximated by field differences divided by the corresponding grid
distances, with weights set by the polar-cell geometry. To include the
vector potential, the difference between neighboring field values is
written as $U_{pq}\Psi_q-\Psi_p$, where
\[
 U_{pq}=\exp\!\left[-i\int_{\bm r_p}^{\bm r_q}
                      \bm A\cdot d\bm l\right].
\]
The integral runs along the grid edge. Under a gauge transformation of
$\Psi$ and $\bm A$, this difference acquires the same local phase as
$\Psi_p$, so its squared modulus is unchanged. This is the
gauge-covariant discretization used here.

We use the symmetric gauge, $A_r=0$ and $A_\varphi=r/r_s^2$.
Radial links are therefore unity. The angular differences are
$U_\varphi(r_i)\Psi_{i,j+1}-\Psi_{i,j}$, with
\begin{equation}
 U_\varphi(r)=\exp\!\left[-i\frac{r^2}{r_s^2}\Delta\varphi\right],
 \qquad \Delta\varphi=\frac{2\pi}{N_\varphi}.
 \label{eq:S_var_link}
\end{equation}
The corresponding angular distance is $r_i\Delta\varphi$.

The interface contributes
\begin{equation}
 F_{B,h}=\frac{r_0\Delta\varphi}{\Gamma}
 \sum_j\left|\Psi_{{\rm out},j}-\Psi_{{\rm in},j}\right|^2.
 \label{eq:S_var_interface_energy}
\end{equation}
No bulk gradient term connects the two sides of the resistive interface.
The resulting numerical energy $F_h$ is used for minimization, energy
comparisons, and stability checks.

\subsection{Vortex states}
\label{subsec:S_variational_multistart}

Several vortex configurations can be locally stable at the same
parameters. To find them, we start from slightly asymmetric centered
vortices, vortices displaced to different positions, and interfacial
soliton profiles. We also gradually increase and decrease $\Gamma$,
using each relaxed solution as the next starting field. An exactly
symmetric starting field can remain centered even when displacement
would lower its energy.

We minimize the numerical free energy $F_h$ using L-BFGS-B~\cite{Byrd1995}, an
iterative method based on energy derivatives with respect to the
field values. These derivatives are calculated automatically from
the same energy expression, using double-precision arithmetic.
The usual target is to reduce the largest derivative component
below $10^{-8}$, with a tighter stopping criterion near transitions.

We identify states from the core position, interface amplitude, and
phase winding. The winding number is the phase change around a closed
contour on which $\Psi$ is nonzero, divided by $2\pi$.
The numbers $N_{\rm in}$, $N_{\rm near}$, and $N_{\rm cell}$ refer
to contours just inside the interface, just outside it, and at the
cell boundary, respectively. In this order, $C$ and $P$ both have
$(1,1,1)$; their cores are centered and displaced, respectively.
The $S$ state has $(0,1,1)$ and no bulk point core.

We test local stability using the second energy derivatives:
small changes of the field must not lower the energy. Changes that
only rotate the configuration can leave the energy unchanged.
Among the locally stable solutions found, the one with the lowest
free energy is taken as the equilibrium state.

\subsection{Locating transition boundaries}
\label{subsec:S_variational_continuation}

We compare stable-state energies to locate $C/P$ and $C/S$ crossings:
\begin{equation}
 F_\alpha(r_0,\Gamma)=F_C(r_0,\Gamma),\qquad \alpha=P,S.
 \label{eq:S_var_centered_crossing}
\end{equation}
The corresponding thresholds are $\Gamma_{CP}$ and $\Gamma_{CS}$.

The local stability of $C$ is tested separately. A small translation contains
angular components $L=0$ and $L=2$,
\begin{equation}
 \delta\Psi_\mu=u_{0,\mu}(r)+u_{2,\mu}(r)e^{2i\varphi},
 \label{eq:S_var_translation_harmonics}
\end{equation}
where $\mu={\rm in,out}$. The energy cost of a small displacement is the
translational curvature $\lambda_{\rm tr}$ of main-text Eq.~(17).
Numerically, we monitor the lowest eigenvalue of the discrete energy
Hessian in this sector. It has the sign of $\lambda_{\rm tr}$, is positive
for local stability, and vanishes at the same $\Gamma_{\rm tr}$. Only this
zero is used to locate thresholds;
Fig.~\ref{fig:S_isolated_cp_order}(a) shows the eigenvalue in this
normalization. This coincides with
$\Gamma_{CP}$ for continuous onset; for a first-order transition it
marks the $C$ spinodal, not the equilibrium crossing.

At $\Gamma_{PS}$, the $P$ core reaches the interface and is absorbed:
\begin{equation}
 (N_{\rm in},N_{\rm near},N_{\rm cell}):
 (1,1,1)\longrightarrow(0,1,1).
 \label{eq:S_var_gammaPS}
\end{equation}
This contact condition differs from energy equality. In the first-order
regime, following $P$ as $\Gamma$ decreases locates the lower $P$ spinodal
$\Gamma_{P,{\rm low}}$, where its local energy minimum disappears.

\subsection{Energy barriers}
\label{subsec:S_variational_constrained}

For the $C/P$ energy profile, we fix the core at $0\le x_c<r_0$ and
locally relax the remaining field:
\begin{equation}
 \mathcal F(x_c;r_0,\Gamma)=
 \min_{\Psi\in\mathcal B_1}\left\{F[\Psi]:\Psi_{\rm in}(x_c,0)=0\right\}.
 \label{eq:S_var_constrained_energy}
\end{equation}
Here $\mathcal B_1$ is the single-vortex family followed from $C$; the
minimization is local within this family. We check the $+1$ winding
around the imposed core and the contour windings, since the point
constraint alone does not fix them. These profiles do not test
exterior-core states. The $x$ axis fixes the displacement direction,
and interpolation imposes the zero between grid points without an
artificial energy term. Figure~\ref{fig:S4_landscapes} shows a minimum
moving away from the center at $r_0=1.5$, and two minima separated by
a barrier at $r_0=2.2$.

After core absorption, a core coordinate cannot describe $S$. We therefore
relax a sequence of fields joining fixed $C$ and $S$ endpoints, keeping
successive fields approximately equally spaced (nudged elastic band/string
method).
This finds a low-energy transition path using the same $F_h$ and its
derivatives. The path is followed as $\Gamma$ changes; its normalized
length is $s$. Figure~\ref{fig:S4_NEB} shows the calculated paths
for $r_0=3.5$.

For film thickness $d$, the physical barrier is
\begin{align}
 E^{\ddagger}_{\rm phys}&=dF_0\Delta F^{\ddagger}\nonumber\\
 &=\frac{2\pi^4}{7\zeta(3)}\,dN(0)(k_BT_c)^2\xi_s^2(1-t)
 \Delta F^{\ddagger},
 \label{eq:S_dimensional_barrier}
\end{align}
where $t=T/T_c$ and $F_0$ is given by Eq.~\eqref{eq:S_energy_unit}.
At fixed dimensionless parameters, this barrier scales with
$d(1-t)$; temperature changes can also change $\Delta F^\ddagger$.

\subsection{Numerical checks}
\label{subsec:S_variational_mesh_tests}

Radial and angular refinement was checked on three successively finer
grids at $r_0=0.25$, $1.5$, and $3.0$, keeping the interface aligned
with the grid. Minimization errors were smaller than the refinement
changes. Numerical differences of $F_h$ checked its calculated
derivatives, and a separate radial calculation confirmed the energy
change for a small displacement. Section~\ref{sec:S_additional_diagnostics}
gives further checks of reverse parameter sweeps, local stability,
angular orientation, and interface current continuity.

\section{Independent check using finite differences}
\label{sec:S_fd}

We also solve the GL equations and interface conditions directly by
finite differences (FD), independently of the discrete energy in
Sec.~\ref{sec:S_variational}. This checks selected stationary fields
and transition thresholds. Unless stated otherwise, $r_s=\sqrt{20}$.

\subsection{Equations and boundary conditions}
\label{subsec:S_fd_continuum}

In each region, $\mu={\rm in,out}$, we solve
\begin{equation}
 -(\bm\nabla-i\bm A)^2\Psi_\mu-\Psi_\mu
 +|\Psi_\mu|^2\Psi_\mu=0,
 \label{eq:S_fd_GL}
\end{equation}
with $\bm A=(r/r_s^2)\bm e_\varphi$ and the interface conditions
\eqref{eq:S_BC_in} and \eqref{eq:S_BC_out}. At the cell boundary,
$\Psi_{\rm out}(r_s,\varphi)=f_s e^{i\varphi}$. A separate centered
radial FD solution with $f'_{\rm out}(r_s)=0$ gives $f_s$, which is held the same for all states
compared at fixed $(r_0,\Gamma,r_s)$.

We choose the displacement axis by solving on $0\le\varphi\le\pi$
with reflection symmetry,
\begin{equation}
 \Psi(r,-\varphi)=\Psi^*(r,\varphi).
 \label{eq:S_fd_reflection}
\end{equation}
Thus $\operatorname{Im}\Psi=0$ and
$\partial_\varphi\operatorname{Re}\Psi=0$ at $\varphi=0,\pi$.
This fixes the orientation, not the core position. Reflection gives
the full-cell fields used for energy and winding calculations.

\subsection{Numerical solution}
\label{subsec:S_fd_mesh}

The polar grids have radial spacings $h_{\rm in}=r_0/N_{r,{\rm in}}$
and $h_{\rm out}=(r_s-r_0)/N_{r,{\rm out}}$, and angular spacing
$\Delta\varphi=2\pi/N_\varphi$. The representative $r_0=1.5$ grid has
18 inner and 36 outer radial intervals, with $N_\varphi=48$ for the
full circle. Second-order differences include the angular phase factor
in Eq.~\eqref{eq:S_var_link}; the interface conditions use second-order
one-sided radial derivatives.

Reflection makes the origin value $\Psi_0$ real. Its Laplacian uses
the angular average $\overline\Psi_1$ on the first radial ring:
\begin{equation}
 \nabla^2\Psi(0)\simeq\frac{4}{h_{\rm in}^2}
 (\overline\Psi_1-\Psi_0).
 \label{eq:S_fd_origin_laplacian}
\end{equation}
The origin value can therefore be zero for $C$ or nonzero for $P$ and $S$.

We solve the nonlinear equations by Newton iteration, reducing each
field update until it lowers the equation residual. Spatial accuracy
is assessed by grid refinement, rather than the
residual alone. As $\Gamma$ changes, the previous solution supplies the
next starting field and the radial calculation updates $f_s$.

\subsection{Locating transitions}
\label{subsec:S_fd_diagnostics}

Linearizing about $C$ gives the lowest translation eigenvalue
$\lambda_{\rm tr}^{\rm FD}$. As in Sec.~\ref{subsec:S_variational_continuation},
only its zero is used:
\begin{equation}
 \lambda_{\rm tr}^{\rm FD}(r_0,\Gamma)=0.
 \label{eq:S_fd_spinodal}
\end{equation}
This condition marks loss of local stability against displacement. We compare this
threshold with the energy-minimization result; the two eigenvalues are
normalized differently, but their zeros agree at the percent level
(Table~\ref{tab:S4_boundary_skeleton}). $P/S$ absorption is identified by the inner-interface zero
and winding change in Eq.~\eqref{eq:S_var_gammaPS}.

We independently integrate the continuum GL energy over the reconstructed
full cell, including the interface term. This uses radial cubic-spline
and angular Fourier derivatives, with finer radial integration.
Energy equality between $C$ and $P$ or $S$ gives $\Gamma_{CP}$ or
$\Gamma_{CS}$. At continuous $C/P$ onset, the very small energy gain can
fall below the integration accuracy; we then use the FD translation
threshold, $\Gamma_{CP}=\Gamma_{\rm tr}$.

\subsection{Comparison and accuracy}
\label{subsec:S_fd_field_comparison}

At $r_0=1.5$, $C$, $P$, and $S$ at $\Gamma=0.4$, $0.9$, and $1.5$ give
relative root-mean-square amplitude differences below $0.1\%$ between
FD and energy minimization on the representative grid. Grid refinement
shows approximately second-order convergence of the fields, core position,
and minimum soliton interface amplitude. Current continuity uses the same
second-order interface derivatives.

For $P$ and $S$, energy-minimization solutions provide only the starting
fields; the final FD fields solve separately discretized equations.
This checks the same stationary solutions, rather than independently
searching for every energy minimum. The energy-minimization calculation
finds the competing stable states and transition barriers.

\section{Transition regimes under strong caging}
\label{sec:S_finite_confinement}

The compact cell $r_s=\sqrt{20}$, with $H/H_{c2}=0.1$, illustrates
three transition regimes. Its outer boundary confines the vortex more
strongly than the $r_s=16$ cell in main-text Fig.~5, shifting the
transition coordinates and barrier heights.

\subsection{Transition boundaries}
\label{subsec:S4_boundaries}

\begin{figure}[htbp]
\centering
\includegraphics[width=0.98\linewidth]{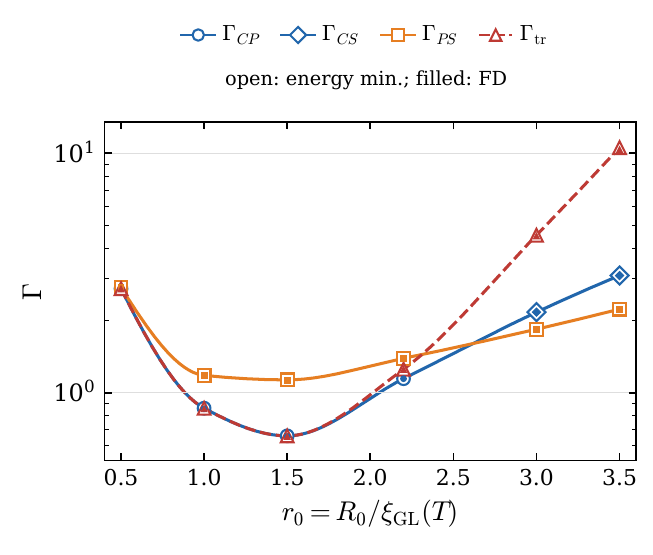}
\caption{Transition boundaries at six radii in the compact cell.
Blue marks the $C/P$ or $C/S$ energy crossing, orange marks $P/S$
conversion, and red dashed curves mark the centered-state stability
limit. Circles, diamonds, squares, and triangles denote $C/P$, $C/S$,
$P/S$, and the stability limit, respectively. Open markers show
energy-minimization results; smaller filled markers show FD results.}
\label{fig:S4_phase_skeleton}
\end{figure}

Table~\ref{tab:S4_boundary_skeleton} and
Fig.~\ref{fig:S4_phase_skeleton} compare energy-minimization and
finite-difference (FD) results at six radii. The methods give the same
boundary ordering, with differences at the percent level.

\begin{table*}[t]
\caption{Transition coordinates in the compact cell. FD values are followed
by energy-minimization values in parentheses. $\Gamma_{CP}$ and
$\Gamma_{CS}$ mark energy crossings, $\Gamma_{\rm tr}$ is the centered
state's translation threshold, and $\Gamma_{PS}$ marks core absorption
as the displaced core reaches the interface.}
\label{tab:S4_boundary_skeleton}
\centering
\footnotesize
\setlength{\tabcolsep}{3.2pt}
\begin{tabular}{cccccc}
\toprule
$r_0$ & $\Gamma_{CP}$ & $\Gamma_{CS}$ & $\Gamma_{PS}$ & $\Gamma_{\rm tr}$
& equilibrium sequence\\
\midrule
0.5 & $2.714$ ($2.719$) & $-$ & $2.746$ ($2.744$)
    & $2.714$ ($2.719$)
    & continuous $C/P$ onset followed by $P/S$ core absorption (narrow $P$)\\
1.0 & $0.856$ ($0.861$) & $-$ & $1.176$ ($1.181$)
    & $0.856$ ($0.861$)
    & continuous $C/P$ onset followed by $P/S$ core absorption\\
1.5 & $0.656$ ($0.660$) & $-$ & $1.127$ ($1.131$)
    & $0.656$ ($0.660$)
    & continuous $C/P$ onset followed by $P/S$ core absorption\\
2.2 & $1.144$ ($1.144$) & $-$ & $1.386$ ($1.392$)
    & $1.251$ ($1.253$)
    & first-order $C/P$ onset followed by $P/S$ core absorption\\
3.0 & $-$ & $2.169$ ($2.169$) & $1.834$ ($1.841$)
    & $4.513$ ($4.548$) & direct $C\to S$\\
3.5 & $-$ & $3.084$ ($3.087$) & $2.225$ ($2.233$)
    & $10.38$ ($10.56$) & direct $C\to S$\\
\bottomrule
\end{tabular}
\end{table*}

At $r_0=0.5$, $1.0$, and $1.5$, continuous $C/P$ onset is followed
by $P/S$ core absorption, with
$\Gamma_{CP}=\Gamma_{\rm tr}<\Gamma_{PS}$.
At $r_0=2.2$, $\Gamma_{CP}<\Gamma_{\rm tr}<\Gamma_{PS}$:
a first-order $C\to P$ transition is followed by $P/S$ core absorption.
At $r_0=3.0$ and $3.5$, $\Gamma_{PS}<\Gamma_{CS}<\Gamma_{\rm tr}$,
so the soliton already exists as a metastable state before the direct
equilibrium $C\to S$ transition.

\subsection{Following the vortex core and the soliton}
\label{subsec:S4_coordinates}

Along the displacement axis, the point core is at $x_c$, where
$\Psi_{\rm in}(x_c,0)=0$. After this zero disappears, we follow the
position $x_{\min}$ and value $m_{\min}$ of the smallest inner amplitude
along $0\le x\le r_0$. The pair $(x_{\min},m_{\min})$ is $(0,0)$ for
$C$, $(x_c,0)$ for $P$, and $(r_0,m_{\min}>0)$ for $S$.

\subsection{\texorpdfstring{Continuous and first-order $C\to P$ transitions}
{Continuous and first-order $C$ to $P$ transitions}}
\label{subsec:S4_CP_order}

\begin{figure}[htbp]
\centering
\includegraphics[width=0.98\linewidth]{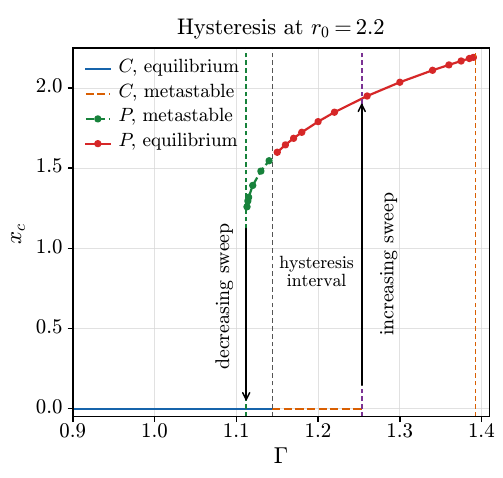}
\caption{Equilibrium and metastable states at $r_0=2.2$ in the compact
cell. The green and purple vertical guides mark the lower $P$ and
centered-state spinodals, which bound the hysteresis interval.
The gray guide marks the $C/P$ energy crossing; the orange guide
marks $P/S$ conversion. Following each local minimum to its spinodal
gives the limiting hysteresis sweep; thermal activation, noise, and
imperfections can cause earlier switching.}
\label{fig:S4_hysteresis}
\end{figure}

At $r_0=2.2$, both $C$ and $P$ are locally stable over
$\Gamma_{P,{\rm low}}<\Gamma<\Gamma_{\rm tr}$, with
$\Gamma_{P,{\rm low}}<\Gamma_{CP}<\Gamma_{\rm tr}$.
This permits hysteresis: as $\Gamma$ increases, $C$ can persist beyond
the energy crossing to its spinodal; as $\Gamma$ decreases, $P$ can
persist to its lower spinodal (Fig.~\ref{fig:S4_hysteresis}). An energy
barrier separates the two states despite the arbitrary displacement
direction.

\begin{figure*}[t]
\centering
\includegraphics[width=0.96\textwidth]{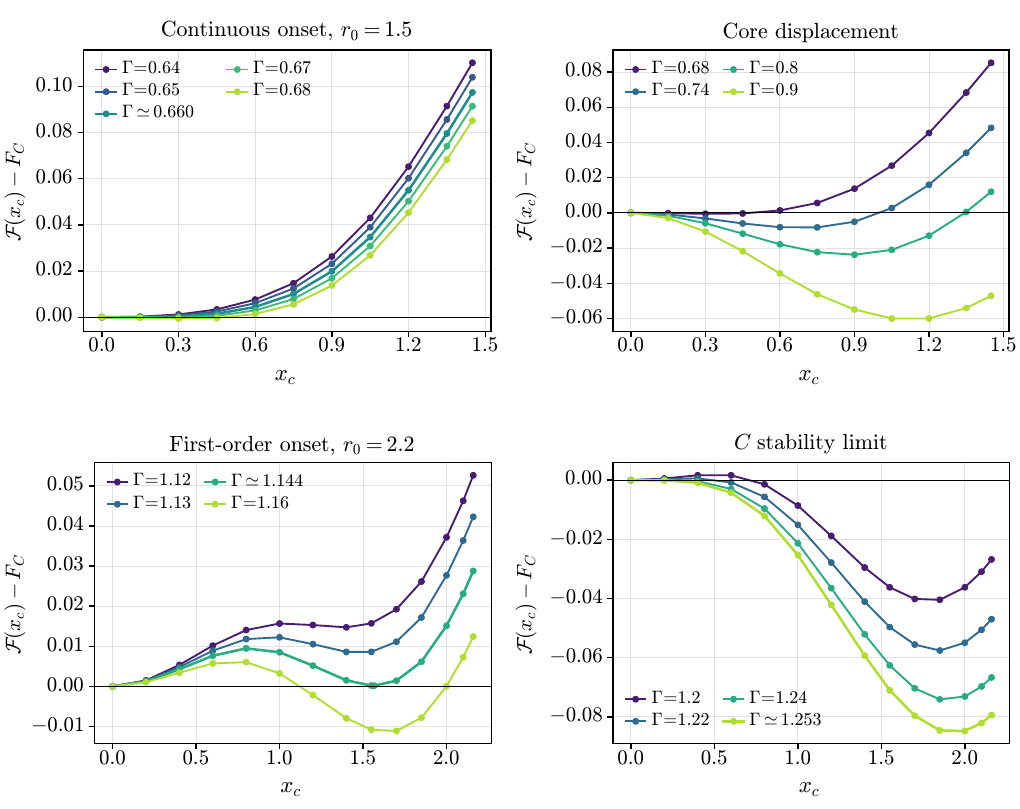}
\caption{Relaxed energy versus core position in the compact cell. Here
$x_c$ is the position of the point core along the displacement axis, and
$\mathcal F(x_c)$ is the free energy after all other field variables are
relaxed at fixed $x_c$ [Eq.~\eqref{eq:S_var_constrained_energy}]. They
correspond to $a$ and $F_{\rm rel}(a)$ in main-text Fig.~6. Top:
continuous onset at $r_0=1.5$, with one moving minimum. Bottom:
first-order onset at $r_0=2.2$, with two minima separated by a barrier.
Points are relaxed solutions at fixed core position; lines guide the eye.}
\label{fig:S4_landscapes}
\end{figure*}

The relaxed energy in Fig.~\ref{fig:S4_landscapes}, calculated at fixed
core position using Eq.~\eqref{eq:S_var_constrained_energy}, explains
the difference between the two onsets. These compact-cell calculations
provide explicit examples of the continuous and first-order profiles
sketched in main-text Fig.~6. At $r_0=1.5$, one minimum moves
continuously away from the center. At $r_0=2.2$, centered and displaced
minima coexist, and their energies cross while a barrier remains.

\subsection{\texorpdfstring{Direct $C\to S$ transition and energy barrier}
{Direct $C$ to $S$ transition and energy barrier}}
\label{subsec:S4_largegrain}

At $r_0=3.5$, the displaced core reaches the interface before the
$C/S$ energy crossing. Its amplitude minimum then stays at the
interface and rises from zero as $S$ forms. The soliton becomes the
equilibrium state at the crossing, while $C$ remains metastable up to
its higher stability limit (Table~\ref{tab:S4_boundary_skeleton}).

\begin{figure*}[t]
\centering
\includegraphics[width=0.98\textwidth]{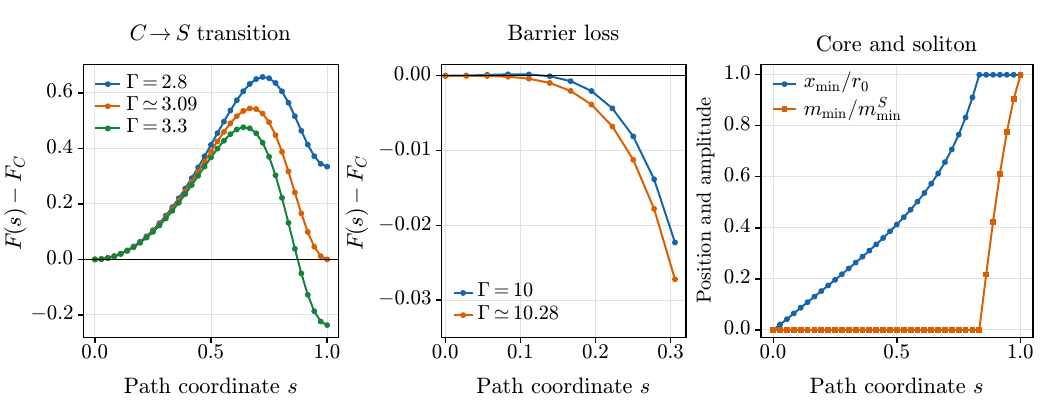}
\caption{Direct $C\to S$ transition paths at $r_0=3.5$.
The path coordinate $s$ is the normalized path length from $C$ ($s=0$)
to $S$ ($s=1$) [Sec.~\ref{subsec:S_variational_constrained}]. It plays
the role of $u$ in main-text Fig.~6.
Left: energy relative to $C$ below, near, and above the equilibrium
crossing. Middle: barrier disappearance on the auxiliary path grid.
Right: along the path near $\Gamma_{CS}$, the position
$x_{\min}/r_0$ and amplitude $m_{\min}/m_{\min}^{S}$ are normalized
by the grain radius and the final soliton amplitude, respectively.
The minimum moves to the interface, then rises from zero as the
soliton forms. Symbols mark calculated configurations; lines join them.}
\label{fig:S4_NEB}
\end{figure*}

The transition paths in Fig.~\ref{fig:S4_NEB} provide calculated
examples of the $C/S$ configuration paths sketched in main-text Fig.~6.
They retain a barrier
$\Delta F^\ddagger\simeq0.54$ near the energy crossing
$\Gamma_{CS}\simeq3.09$. The barrier decreases toward the centered
state's spinodal. On the coarser auxiliary path grid it disappears near
$\Gamma\simeq10.28$ (Fig.~\ref{fig:S4_NEB}, middle), within 3\% of the
refined stability thresholds in Table~\ref{tab:S4_boundary_skeleton}, which are
calculated separately.

\subsection{Interface structure of the soliton}
\label{subsec:S4_phase_profile}

\begin{figure*}[t]
\centering
\includegraphics[width=0.94\textwidth]{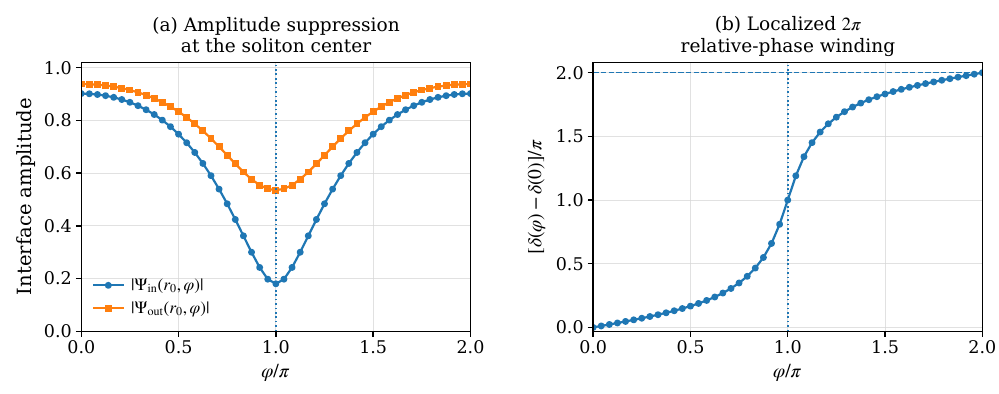}
\caption{Interface structure of $S$ at $r_0=1.5$ and $\Gamma=1.5$.
(a) Inner and outer amplitudes; the dotted line marks the soliton center.
(b) Relative phase measured continuously from its value at
$\varphi=0$, with a total increase of $2\pi$.}
\label{fig:S4_phase_profile}
\end{figure*}

Figure~\ref{fig:S4_phase_profile} shows the interface fields at
$r_0=1.5$ and $\Gamma=1.5$. The relative phase
\begin{equation}
 \delta(\varphi)=\arg\Psi_{\rm out}(r_0,\varphi)
 -\arg\Psi_{\rm in}(r_0,\varphi)
 \label{eq:S4_relative_phase}
\end{equation}
is followed continuously around the interface. Its total change is
$2\pi$, concentrated near the inner-amplitude minimum. The amplitude
remains nonzero there and no point core is present, identifying the
interfacial soliton.

\section{Additional convergence and stability tests}
\label{sec:S_additional_diagnostics}

The following tests check the compact-cell results in
Sec.~\ref{sec:S_finite_confinement}. The larger $r_s=16$ cell is
discussed separately in Sec.~\ref{sec:S_finite_size}.

\subsection{Reversibility of $P/S$ conversion}

At $r_0=1.5$ and $r_s=\sqrt{20}$, continuation through $P/S$
in both directions gives the same energies within numerical accuracy.
Along these configurations, $F$ and $\partial F/\partial\Gamma$ are
continuous across $\Gamma_{PS}$ within numerical accuracy. The derivative
includes the $\Gamma$ dependence of the reference edge amplitude
$f_s(\Gamma)$. Radial and angular refinement preserves this result.
The inner-interface amplitude evolves continuously as the point core
reaches the interface and disappears. These checks support smooth
core absorption without $P/S$ hysteresis in the tested case.

\subsection{Grid convergence of the translation threshold}

The translation stability calculation was repeated with finer radial
and angular grids at $r_0=0.25$, $1.5$, and $3.0$.
Extrapolation in the squared grid spacing gave converged translation
thresholds in these representative regimes. A refined calculation at
$r_0=0.25$ gives $\Gamma_{\rm tr}\simeq6.84946$ and a positive relaxed
quartic coefficient. Constrained energy profiles support a continuous
$C/P$ onset and resolve a narrow $P$ interval,
$\Gamma_{PS}-\Gamma_{\rm tr}\simeq9.8\times10^{-4}$.
The translation threshold tests local stability; energy comparisons
are needed to determine the equilibrium transition.

\subsection{Interface-current convergence}

The discrete interface energy enforces current matching. To check
spatial accuracy, we also reconstruct the inner and outer normal
currents from one-sided radial derivatives. Their largest pointwise
difference decreases with radial refinement for the tested $P$ and
$S$ states. The small-grain case converges more slowly because the
radial scales inside and outside the grain differ strongly.

\subsection{Local stability and angular-grid effects}

To test local stability, we evaluated the second derivative of the free energy
for representative $P$ and $S$ solutions, with respect to small perturbations of
the order-parameter fields.
Rotating the configuration leaves the energy unchanged
in a circular cell; the corresponding calculated curvature was zero
within numerical accuracy. The remaining calculated low-lying modes
had nonnegative curvature, supporting local stability of these states.
Starting the core or soliton on an angular grid line or halfway between
two lines gave the same energy within numerical accuracy. Thus the
angular grid did not select a significant preferred orientation in
the tested cases.

\clearpage
\onecolumngrid
\begin{center}
\centering
\includegraphics[width=0.94\textwidth]{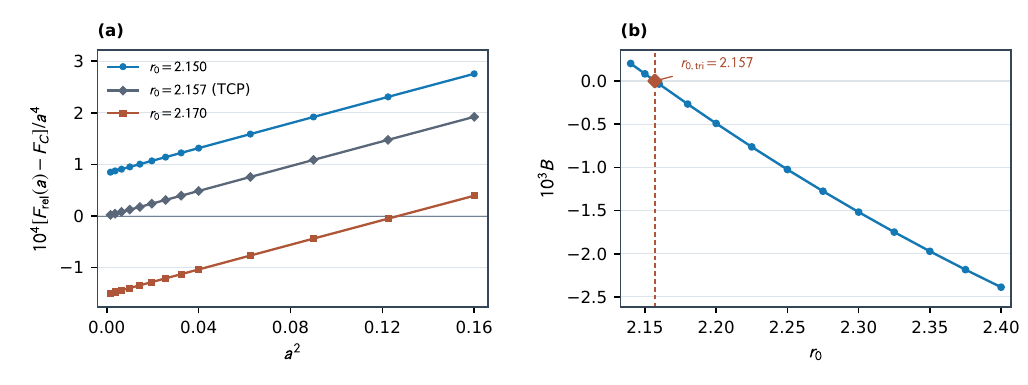}
\captionof{figure}{Landau analysis in the $r_s=16$ cell.
(a) Relaxed energy profiles along $\Gamma_{\rm tr}$ below, near,
and above the tricritical radius. At $A=0$, the intercept and initial
slope of $(F_{\rm rel}-F_C)/a^4$ versus $a^2$ give $B$ and $D$
after removing the displayed axis scale factor.
(b) The quartic coefficient $B$ changes sign at the tricritical
point, marked by the diamond. Symbols show calculated values;
lines guide the eye.}
\label{fig:S_rs16_landau}
\end{center}
\twocolumngrid

\section{Cell-size dependence of vortex-lattice caging}
\label{sec:S_finite_size}

Main-text Fig.~5 uses $r_s=16$. To locate its tricritical point,
we expand the energy after relaxing the field at fixed core
displacement $a$~\cite{LandauLifshitz1980}:
\[
 F_{\rm rel}(a)-F_C=Aa^2+Ba^4+Da^6+O(a^8).
\]
Rotational symmetry permits only even powers. The centered-state
stability limit has $A=0$. Near the change of order, $D>0$:
positive $B$ gives continuous $C/P$ onset, while negative $B$ gives
a first-order transition. The tricritical point has $A=B=0$ and
$D>0$. The calculation gives
$(r_{0,\rm tri},\Gamma_{\rm tri})\simeq(2.157,0.1167)$.
Figure~\ref{fig:S_rs16_landau} directly supports the tricritical
profile in main-text Fig.~6(b). Along $\Gamma_{\rm tr}$,
$(F_{\rm rel}-F_C)/a^4=B+Da^2+O(a^4)$, so the intercept and initial
slope give $B$ and $D$. At tricriticality the intercept vanishes
while the slope stays positive. Numerical profiles also confirm
$a_{\min}\propto(\Gamma-\Gamma_{\rm tr})^{1/4}$.

At first-order $C/P$ coexistence, the two minima have equal energies.
At the lower $P$ spinodal, the displaced minimum and saddle merge:
$\partial_aF_{\rm rel}=\partial_a^2F_{\rm rel}=0$. The thresholds obey
$\Gamma_{P,{\rm low}}<\Gamma_{CP}<\Gamma_{\rm tr}$.
During slow sweeps, $C$ can persist up to $\Gamma_{\rm tr}$ and
$P$ down to $\Gamma_{P,{\rm low}}$, producing hysteresis.
The displacement jump and both metastability intervals vanish
at the tricritical point. These conditions were checked by constrained
relaxation (Sec.~\ref{subsec:S_variational_constrained}), with matched
grids and radial and angular refinement to resolve the small threshold
separations.

The equilibrium $P$ region ends where continuous core absorption
meets the $C/P$ energy crossing, near $(r_0,\Gamma)\simeq(5.37,0.732)$.
This junction, distinct from the tricritical point, marks the change
to a direct first-order equilibrium $C\to S$ transition. Cell size changes this sequence:
at $r_0=3$, the compact $r_s=\sqrt{20}$ cell has a direct $C\to S$
transition (Table~\ref{tab:S4_boundary_skeleton}), while $r_s=16$
retains an equilibrium $P$ interval. Weaker confinement therefore
extends the $P$ region and reduces the interval in which $C$
remains metastable at this radius. The $P/S$ line marks morphological core absorption, with continuously
evolving fields in the tested cases.

\FloatBarrier